\documentclass[aps,pre,floatfix,longbibliography]{revtex4-1}
\usepackage{amsfonts}
\usepackage{amsmath}
\usepackage{amsthm}
\usepackage[x11names]{xcolor}
\usepackage{graphics}
\usepackage{graphicx,epstopdf}
\usepackage[export]{adjustbox}
\usepackage{subfigure}
\usepackage{amssymb}
\usepackage{lineno}
\usepackage[colorinlistoftodos]{todonotes} 
\usepackage[T1]{fontenc}
\usepackage{amsmath,amssymb,epsfig,latexsym,natbib,graphicx,dcolumn}

\colorlet{BLUE}{blue}
\colorlet{RED}{red}
\colorlet{green}{Green3}

\begin{document}
\setcounter{page}{1}
\title{Non-Boussinesq subgrid-scale model with dynamic tensorial coefficients }
\author{Rahul Agrawal}\email{rahul29@stanford.edu}
\affiliation{Center for Turbulence Research, Stanford University,  California, United States of America-94305}

\author{Michael P. Whitmore}
\affiliation{Center for Turbulence Research, Stanford University,  California, United States of America-94305}
\author{Kevin P. Griffin}
\affiliation{Center for Turbulence Research, Stanford University,  California, United States of America-94305}
\author{Sanjeeb T. Bose}
\affiliation{Cascade Technologies, Inc., California, United States of America-94303 and Institute for Computational and Mathematical Engineering, Stanford University}
\author{Parviz Moin}\email{moin@stanford.edu}
\affiliation{Center for Turbulence Research, Stanford University,  California, United States of America-94305}
\date{\today}
 \begin{abstract}

A major drawback of Boussinesq-type subgrid-scale stress models used in large-eddy simulations is the inherent assumption of alignment between large-scale strain rates and filtered subgrid-stresses. {\emph{ A priori}} analyses using direct numerical simulation (DNS) data  has shown that this assumption is invalid locally as  subgrid-scale stresses are poorly correlated with the large-scale strain rates [\textit{Bardina et al., AIAA 1980; Meneveau and Liu, Ann. Rev. Fluid Mech. 2002}]. In the present work, a new, non-Boussinesq subgrid-scale model is presented where the model coefficients are computed dynamically. 
Some previous non-Boussinesq models have observed issues in providing adequate dissipation of turbulent kinetic energy [e.g.: \textit{Bardina et al., AIAA 1980; Clark et al. J. Fluid Mech., 1979; Stolz and Adams, Phys. of Fluids, 1999}]; however, the present model is shown to provide sufficient dissipation using dynamic coefficients.
Modeled subgrid-scale Reynolds stresses satisfy the consistency requirements of the governing equations for LES, vanish in laminar flow and at solid boundaries, and have the correct asymptotic behavior in the near-wall region of a turbulent boundary layer. 

The new model, referred to as the dynamic tensor-coefficient Smagorinsky model (DTCSM), has been tested in  simulations of canonical flows: decaying and forced homogeneous isotropic turbulence (HIT), and wall-modeled turbulent channel flow at high Reynolds numbers; the results show favorable agreement with DNS data. It has been shown that DTCSM offers similar predictive capabilities as the dynamic Smagorinsky model for canonical flows. In order to assess the performance of DTCSM in more complex flows, wall-modeled simulations of high Reynolds number flow over a Gaussian bump (Boeing Speed Bump) exhibiting smooth-body flow  separation are performed. Predictions of surface pressure and skin friction, compared against DNS and experimental data, show improved accuracy from DTCSM in comparison to 
existing static coefficient (Vreman) and dynamic Smagorinsky model. \textcolor{black}{The computational cost of performing LES with this model is up to 15\% higher than the dynamic Smagorinsky model.} 


\end{abstract}
\maketitle

\section{Introduction}


The most commonly used class of subgrid-scale stress models is the eddy-viscosity formulation. Smagorinsky \cite{smagorinsky1963general} developed a subgrid-scale (SGS) stress closure model \textcolor{black}{in which} the subgrid-scale stress are assumed to be aligned and scaled with the local 
strain rate of the large-scale eddies.  This model produces satisfactory results in simulations of decaying homogeneous isotropic turbulence \cite{lilly1966application,mansour1979improved}; however, the modeled SGS stress does not vanish in laminar regions nor near solid walls, making it unsuitable for simulating transitional or wall-bounded flows. To account for near-wall scalings, the model was modified by prescribing damping functions \cite{moin1982numerical,piomelli1988model} that return the correct asymptotic behavior near solid boundaries. Germano et al. \cite{germano1991dynamic} developed the dynamic Smagorinsky model (DSM) in which the model coefficient is dynamically computed from the local flow state without any prescribed coefficients, improving the predictive capability of LES. The dynamic variant of the Smagorinsky model was shown to accurately dissipate energy from the large-scales in simulations of isotropic decaying turbulent flow, and appropriately vanish in laminar and transitional flows \cite{piomelli1993high}. However, to ensure numerical stability, a regularization procedure, such as clipping or spatial averaging, is often required to be applied to the dynamically computed model coefficient \citep{ghosal1995dynamic}.  

To avoid challenges that arise in complex flows and on unstructured grids, such as the construction of test filters or the definition of homogeneous averaging operations, more sophisticated static-coefficient 
models have been proposed. These models attempt to embed properties such as appropriate asymptotic 
near-wall scaling \citep{nicoud_wale}, vanishing eddy viscosities in laminar regions 
\citep{vreman2004eddy,nicoud_sigma}, or minimum\textcolor{black}{, but, sufficient} dissipation of small-scale 
turbulence \citep{rozema2015minimum}.  While these models have reduced computational 
complexity compared to their dynamic counterparts, they have been seen to have relatively weaker predictive capability \textcolor{black}{in complex flows}. For  instance, recent simulations of flows around realistic aircraft models have shown 
that the dynamic Smagorinsky model offers more accurate predictions of integrated 
quantities of interest (e.g., lift, drag) and salient flow features (e.g., separation 
bubble extents) \citep{gocsubgrid}.  

Lastly, most investigations into the construction of subgrid-scale models for large-eddy 
simulation have focused on their performance in homogeneous isotropic turbulence and wall resolved LES limits. The resolution 
of viscously scaled eddies near the wall is prohibitively expensive at high Reynolds 
numbers and recent advances have shown the relatively successful application of wall 
modeled LES approaches to practical engineering flows \citep{bose2018wall}.  In many of 
these wall modeled LES calculations, detailed comparisons of different subgrid-scale models 
have not been available \citep{goc2020wall}. 


 This investigation proposes a novel, dynamic subgrid-scale model that \textcolor{black}{ does not assume alignment between the subgrid stresses and resolved strain rates through the introduction of a tensorial eddy viscosity in contrast to traditional isotropic eddy viscosity closures}. It is shown that this tensorial eddy viscosity better correlates with  the local subgrid stress (in \emph{a priori} tests) and offers similar accuracy to existing dynamic models in isotropic limits (homogeneous turbulence). Detailed  \emph{a posteriori} testing of this model is focused on high Reynolds number limits where interactions of SGS and wall models have to be considered. For this, we discuss the performance of the proposed model in both canonical (turbulent channel flow) and complex flow exhibiting separation.  In these wall modeled LES calculations, the dynamic tensorial  SGS \textcolor{black}{model outperforms} existing static and dynamic coefficient models.

This paper is organized as follows. We revisit the existing SGS formulations in Section II, and then present the proposed modeling approach in Section III. The details of the various numerical solvers used in this \textcolor{black}{work} are provided in Section IV. \emph{A priori} results from these models including the stress tensor and kinetic energy dissipation based correlations between modeled and exact stresses for turbulent channel flow are presented in Section V. Asymptotic behavior of the proposed model near a solid wall and in laminar channel flow are discussed in Section VI. Detailed \emph{a posteriori} analysis of the performance of these models in isotropic turbulence and high Reynolds number channel flows is presented in Section VII. The model is subsequently applied to a flow over a Guassian bump exhibiting separation and comparisons of the surface pressure and skin friction between WMLES and spanwise-periodic quasi-DNS and 3D experiments are presented in Section VIII. \textcolor{black}{Remarks about the costs incurred on using DTCSM in LES are made in Section IX. Conclusions are offered in Section X. }

\section{LES formalism and governing equations}

In LES, the large-scale quantities are defined by filtering the velocity and pressure fields. If the grid-filter kernel operator is denoted by $\mathcal{G}$, then a large scale quantity, $\overline{f}$ is evaluated from the total field, $f$ as, 
\begin{equation}
\overline{f}(x)    = \int \mathcal{G}(x,x') f(x') dx'
\end{equation}
where the integral is extended \textcolor{black}{over} the entire computational domain. Further, we assume that the grid filter is such that it commutes with the differentiation operation. More details on LES formalism can be found in previous studies \cite{germano1991dynamic,vreman2004eddy,ghosal1995dynamic}.  

The  governing equations for LES of incompressible turbulent flows (of constant density $\rho$) are obtained by applying the aforementioned filter to the Navier-Stokes equations. The resulting equations are
\begin{equation}
    \frac{\partial \overline{u}_i}{\partial x_i} = 0  
\end{equation}
\begin{center}
    and
\end{center}
\begin{equation}
    \frac{\partial \overline{u}_i}{\partial t }+\frac{\partial \overline{u}_j \;  \overline{u}_i}{\partial x_j } =-\frac{1}{\rho}\frac{\partial \overline{p}}{\partial x_i } + \nu \frac {\partial^2  \overline{u}_i}{\partial x_j \partial x_j }-\frac{\partial \tau^{sgs}_{ij}}{\partial x_j}~,
\end{equation}
where $\tau^{sgs}_{ij} = \overline{u_i u_j} - \overline{u}_j \;  \overline{u}_i $ is the subgrid stress which requires modeling closure. The isotropic component of the SGS stress is often absorbed into pressure, which leads to a pseudo-pressure field ($p \mapsto p + 
\rho\tau^{sgs}_{kk}$). The eddy viscosity based SGS closure models based on the 
Boussinesq hypothesis take the form
\begin{equation}
    \tau^{sgs}_{ij} - \frac{1}{3} \tau_{kk} \delta_{ij} = -2 \nu_t \overline{S}_{ij}~,
    \quad {\rm where} \quad
    \overline{S}_{ij} = \frac{1}{2}\left(\frac{\partial \overline{u}_i}{\partial x_j } + \frac{\partial \overline{u}_j}{\partial x_i }\right)~.
\end{equation}
where $\delta_{ij}$ is the Kronecker-delta function.
Two of the more commonly used Boussinesq models that are examined in this \textcolor{black}{work} are the Smagorinsky model with its dynamic variant and the Vreman model, which are described in detail below.

\subsection{Smagorinsky model}

The units of eddy viscosity are a velocity times a characteristic length scale.
Smagorinsky's \cite{smagorinsky1963general} celebrated eddy viscosity model is based on a length scale proportional to the LES grid scale and a velocity scale obtained from the 
product of the grid scale and the magnitude of the strain-rate tensor. With these assumptions, the eddy viscosity is given as
\begin{equation}
    \nu_t = (C_s \Delta)^2 |S|~,\quad{\rm where}\quad|S| = \sqrt{2 \overline{S}_{ij} \overline{S}_{ij}} 
\end{equation}
and $\Delta$ is the grid filter width. Later, Lilly \cite{lilly1970ncar} showed that for isotropic turbulence with spatial resolution that lies in the inertial subrange, $C_s \sim 0.17$.  Deardorff \cite{deardorff1970numerical} recommended $C_s \sim 0.1$ in turbulent shear flows. The applicability of this model is limited, especially in the near-wall region of wall-bounded flows in part because eddy viscosity does not vanish at the wall. Consequently, previous studies \cite{moin1982numerical} have used wall-damping functions \cite{van1956turbulent} to correct for the behavior of eddy viscosity in the viscous sublayer.

\subsection{Dynamic Smagorinsky model}

Germano et al. \cite{germano1991dynamic} introduced the notion of test filtering of the LES governing equations. Through the use of these ideas, the \textcolor{black}{resolved turbulent stresses} \cite{leonard1975energy}, $L_{ij} = - \widehat{\overline{u}_i\overline{u}_j} + \widehat{\overline{u}_i} \; \widehat{\overline{u}_j}$ ($\widehat{(\cdot)}$ denotes test-filter operation) can be related to the \textcolor{black}{modeled stresses in the ``test window'' \cite{lilly1992proposed} } as
\begin{equation}
    L_{ij} = 2\left(C_s \Delta \right)^2 \left( {\frac{\widehat{\Delta}^2}{\Delta^2}} \widehat{|S|}\widehat{S_{ij}} - \widehat{|S|S_{ij}}\right) = 2\left(C_s \Delta \right)^2 M_{ij},
    \label{eqn:dsm}
\end{equation}
where $\widehat{\Delta}$ and $\Delta$ denote test-level and grid-level filter widths, respectively. For an incompressible flow, Eq. (\ref{eqn:dsm}) is an over-determined system with five independent equations for one undetermined coefficient. Lilly  \cite{lilly1992proposed} proposed a least-squares solution of this system, leading to the expression for the model coefficient
\begin{equation}
    (C_s \Delta)^2 =  \frac{L_{ij} M_{ij}}{2 M_{ij} M_{ij}}.
\end{equation}
To avoid numerical instabilities arising from the computation of negative eddy 
viscosities ($C_s < 0$), the numerator and denominator of this equation are averaged in statistically homogeneous directions (and possibly time for statistically stationary flows) to give its working form 
\begin{equation}
     (C_s \Delta)^2 =  \frac{\langle L_{ij} M_{ij}\rangle}{\langle 2 M_{ij} M_{ij}\rangle}, 
\end{equation}
where $\langle \cdot \rangle$ is the spatio-temporal averaging operator. \textcolor{black}{It should be noted that Eq. \ref{eqn:dsm} requires an assumption that $C_s$ can be extracted out of the test filtering operation, or $\widehat{C_s\Delta^2|S|S_{ij}} = C_s \Delta^2  \widehat{|S| S_{ij}}$, which is strictly true only when a regularization procedure, such as spatial averaging (in homogeneous directions) and/or temporal averaging (for statistically stationary flows), is applied \cite{ghosal1995dynamic,meneveau1996lagrangian}.}

\subsection{Vreman model}\label{section:vreman}

Vreman \cite{vreman2004eddy} proposed a modified subgrid-scale model by including the velocity gradient tensor and the gradient model expansion of the exact SGS tensor, leading to the model form
\begin{equation}
    \nu_t = 2.5 (C_s)^2 \sqrt{\frac{B_{\beta}}{\alpha_{ij}\alpha_{ij}}} ~;
    \quad \alpha_{ij} = \frac{\partial \overline{u}_i }{\partial x_j}  ~; \quad \beta_{ij} = \Delta^2 \alpha_{mi}\alpha_{mj} ~; 
\end{equation}
\begin{equation}
    B_{\beta} = \beta_{11}\beta_{22} - \beta_{12}^2 + \beta_{11}\beta_{33} - \beta_{13}^2 + \beta_{22}\beta_{33} - \beta_{23}^2 
\end{equation}
The nominally accepted value of $C_s \approx 0.17$ is used in the present implementation of this model. Unlike the Smagorinsky model, the Vreman model does not over-predict subgrid dissipation in transitional flows, and also appropriately provides zero subgrid dissipation in laminar flows (in two velocity component base states). 

\section{Modeling framework}

While the dynamic procedure improves the predictive capability of the constant coefficient Smagorinsky model, it does not resolve the model form error that is inherent to all Boussinesq SGS models (i.e. that the SGS stress is not necessarily aligned with the strain-rate tensor). Additionally, for LES of flows with mean anisotropy, the assumption of a scalar model coefficient, as used in classical Boussinesq eddy viscosity closures, is potentially overly restrictive. In this work, we propose a novel dynamic formulation for the tensor-coefficient Smagorinsky model \cite{moin1993new}. This model contains non-Boussinesq terms that do not lead to dissipation but would potentially improve the local alignment between modeled and exact subgrid stresses. 
Similar to the dynamic Smagorinsky model, the only input parameter in this model is the ratio of test-level to grid-level filter widths, which is chosen to be two as in previous studies.

\subsection{Dynamic tensor-coefficient Smagorinsky model (DTCSM)}

Moin \cite{moin1993new} proposed the following tensor-coefficient-based Smagorinsky model \textcolor{black}{which alleviates the assumption of alignment between filtered resolved stresses and mean strain rates, }
\begin{equation}
    \tau_{ij}^{sgs} - \frac{\tau^{sgs}_{kk}  }{3} \delta_{ij} = - (C_{ik}S_{kj} + C_{jk}S_{ki} )|S|\Delta^2. 
\label{eqn:dtcsm0}
\end{equation}
This model contains nine independent coefficients, thus providing more degrees of freedom in determining alignment of the stress and strain-rate tensors. However, in its current form, the tracelessness of the model, which is a requirement imposed by the \textcolor{black}{the fact that only the deviatoric part of the exact subgrid stress is being modeled}, is not guaranteed. To treat this, we put the following constraints on the coefficients 
(see Appendix \ref{app:1} for details):
\begin{equation}
    C_{11} = C_{22} = C_{33} ~; \qquad
    C_{ij} = - C_{ji} \quad (j \neq i) ~.    
    \label{eqn:dtcsm1}
\end{equation}
%
The realizability constraints reduce the number of independent coefficients from nine to four. Incorporating this reduction, we invoke the Germano identity to arrive at
\begin{equation}
    L_{ij} = ( C_{ik} \Delta^2 M_{kj} + C_{jk} \Delta^2 M_{ki} ) .
    \label{eqn:dyndtcsm}
\end{equation}
\textcolor{black}{Note that this equation, similar to Eq. \ref{eqn:dsm}, implies taking the model coefficients outside of the test filter, and so the aforementioned regularization procedure is also used in the present case.} For an incompressible flow, this system of five independent equations with four coefficients is solved using the least-squares solution method \cite{lilly1992proposed} to obtain the coefficients dynamically. 
It is noteworthy that only $C_{11}$ (and equivalently $C_{22},C_{33}$) contributes to the dissipation of energy from the large scales.

The reader is directed to Appendices \ref{app:2} and \ref{app:3} for more details on the dynamic procedure and a brief discussion of how the model form can be expressed explicitly in terms of a combination of strain-rate and rotation-rate tensors.

\section{Numerical solver details}

In this paper, two second-order finite-difference solvers with staggered Cartesian meshes are used to simulate homogeneous isotropic turbulence and turbulent channel flow, respectively. These solvers have been validated in previous studies of homogeneous isotropic turbulence \cite{pouransari2016parallel,bassenne2016constant} and turbulent channel flow \cite{lozano2019error,bae2021effect}. Both employ explicit fourth and third-order Runge-Kutta time integration schemes, respectively.

Simulations of the Gaussian bump are performed using a low-dissipation, explicit, unstructured, finite-volume solver for the compressible Navier-Stokes equations  (charLES).  This code is formally 2\textsuperscript{nd}-order accurate in space and 3\textsuperscript{rd}-order accurate in time. More details of the solver and validation cases in subsonic, transonic, and supersonic studies can be found in  \cite{bres2018large,fu2021shock,lakebrink2019toward,goc2021large}. \textcolor{black}{Details of the choice of the computational grid for simulating the Gaussian bump are presented in Section \ref{sec:bump}.}

\section{{A priori} tests}

In this section, we compare \emph{a priori} performance of DSM and DTCSM at stress tensor and dissipation rate levels with respect to filtered-DNS data (a discretely commutative filter with three vanishing moments \cite{vasilyev1998general} is recursively applied to obtain a filter width equal to three times the grid size) for a \textcolor{black}{turbulent channel flow at both low and high friction Reynolds numbers, specifically at $\textit{Re}_{\tau} = 395$ \cite{moser1999direct} and $\textit{Re}_{\tau} = 2000$ \cite{lozano2014effect}. }In the presence of flow anisotropy, an improved SGS model is expected to produce higher stress tensor correlations and provide more realistic representation of the small scales.  These correlations are defined as
\begin{equation}
    \rho = \frac{1}{6}\sum^6_{k=1} \frac{ cov( \tau_{model, k},  \tau_{exact,k} ) }{\sigma( \tau_{models, k}) \sigma( \tau_{exact,k})  },
\end{equation}
where $k = {1,\,2,...,6} $ are the six components of exact ($\tau_{exact}$) and modeled ($\tau_{model}$) SGS stress tensors. Note that $cov(X,Y)$ denotes the covariance between quantities $X$ and $Y$ and $\sigma(X)$ is the standard deviation of the distribution of the quantity $X$. The correlations based on the kinetic energy dissipation rate are similarly defined by contracting the stress tensor with the large-scale strain-rate tensor. 

\begin{figure}[!ht]
    \centering
    \subfigure[]{
       \includegraphics[width=0.45\textwidth]{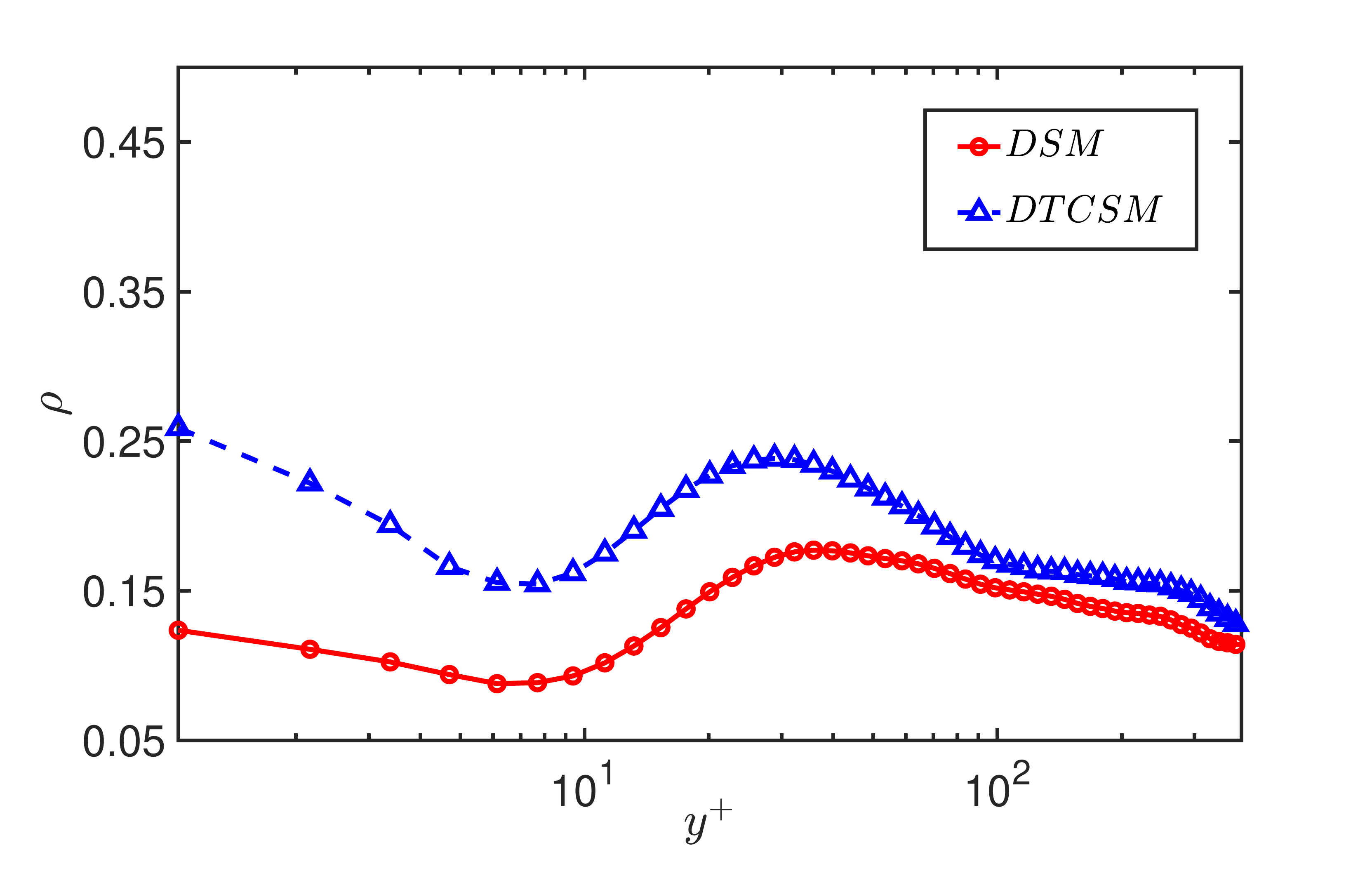}    }
    \subfigure[]{
    \includegraphics[width=0.45\textwidth]{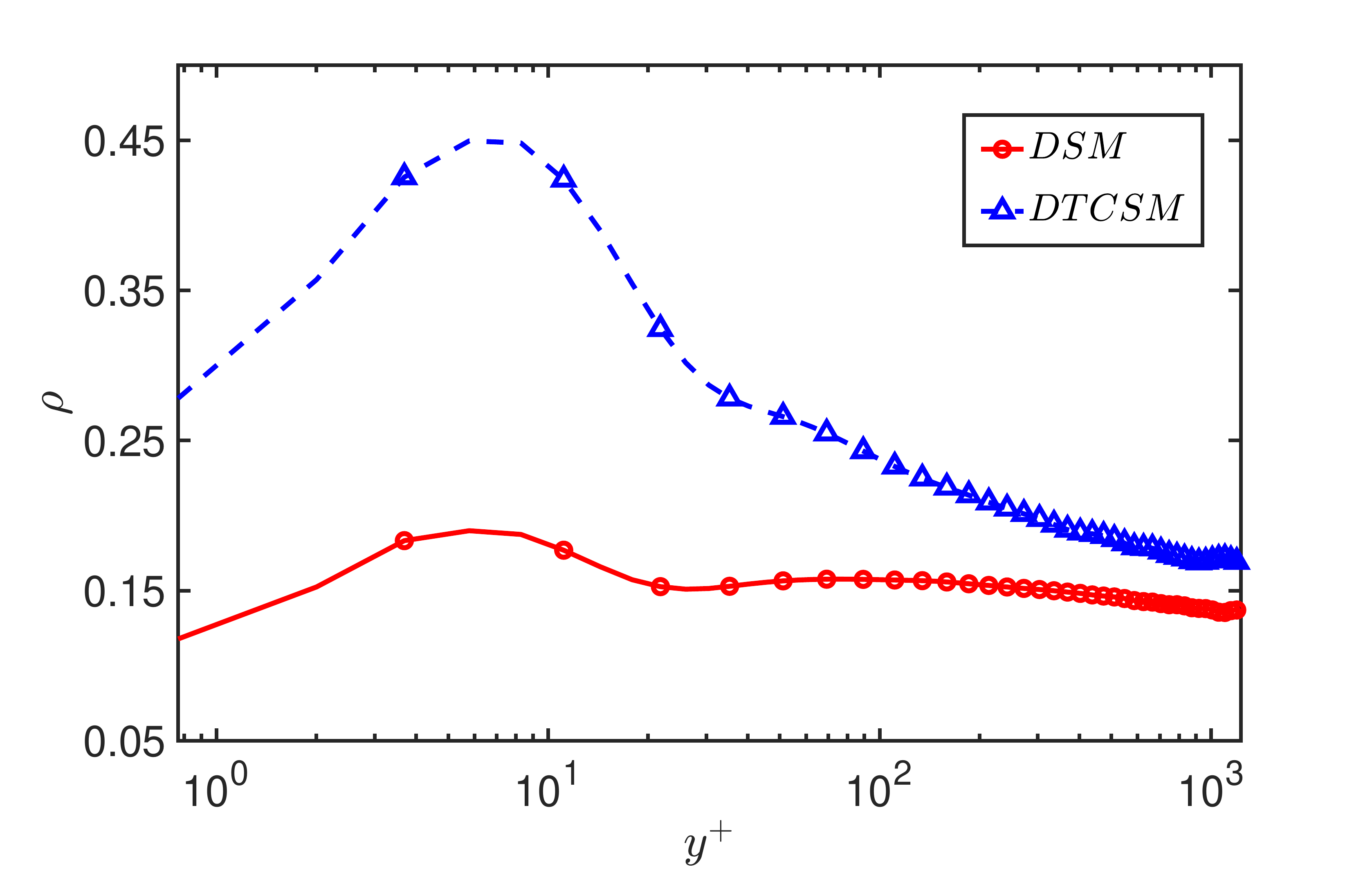}}
    \caption{\color{black} Stress tensor correlations of DSM and DTCSM with respect to filtered DNS of turbulent channel flow at (a) $Re_{\tau} = 395$ and (b) $Re_{\tau}=2000$ respectively. Note that every third value has been marked for the higher Reynolds number case for visual clarity. }
    \label{fig:aprioritensor}
\end{figure}

\textcolor{black}{Figure \ref{fig:aprioritensor} shows that DTCSM produces higher correlations than the dynamic Smagorinsky model at the stress level for both cases. Further, the model performance improves as the Reynolds number is increased from $Re_{\tau} = 395$ to $Re_{\tau} = 2000$. It is also encouraging that the correlations are most improved near the wall, which is the region of high anisotropy and, in practical calculations, the limiting region in terms of resolution requirements}.  {\emph{A priori}} \textcolor{black}{tests}, however, are not \textcolor{black}{necessarily} reflective of the model performance in simulations. The model of Bardina et al. \cite{bardina1980improved} showed much improved tensor-level correlations than even DTCSM; however, it was shown to be significantly under-dissipative in simulations and required ad hoc supplementary dissipation from an eddy viscosity of Smagorinsky form. \textcolor{black}{Although a mixed form of Bardina's scale-similarity model exists \cite{vreman1994formulation}, its application to complex wall-bounded turbulent flows at high Reynolds numbers has been limited. For these reasons, the current analysis has been restricted to comparisons against eddy-viscosity type SGS model formulations. Finally, it should be noted that these correlations are not necessarily a measure of the alignment between modeled and exact subgrid stresses. For more details, the reader is referred to the previous works \cite{higgins2003alignment, tao2002statistical,yang2016topology} where the alignment of subgrid stresses and strain rate tensors are examined through their eigenvalues and eigenvectors. } 

\begin{figure}[!ht]
    \centering
    \subfigure[]{
    \includegraphics[width=0.45\textwidth]{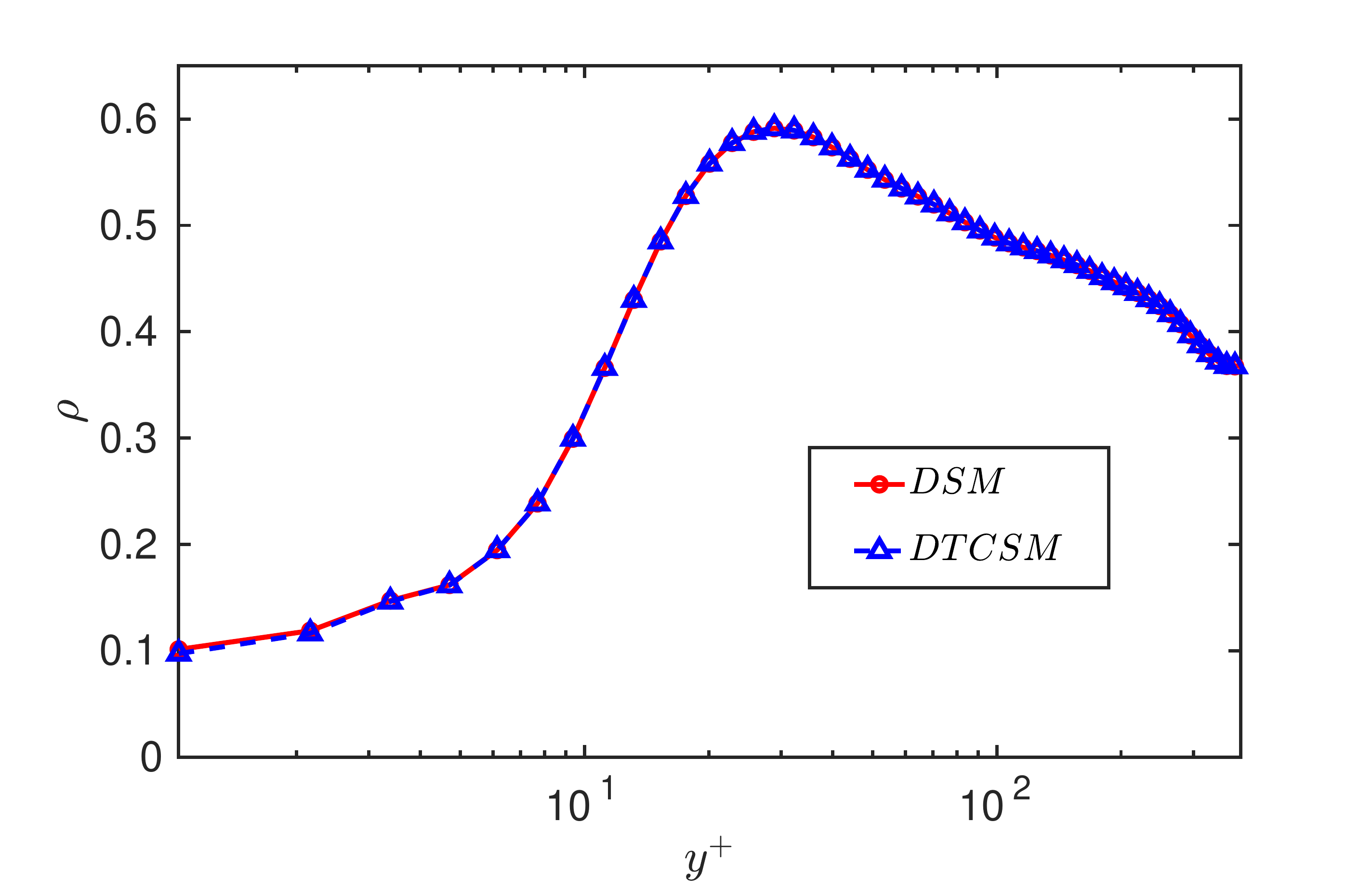}}
    \subfigure[]{
    \includegraphics[width=0.45\textwidth]{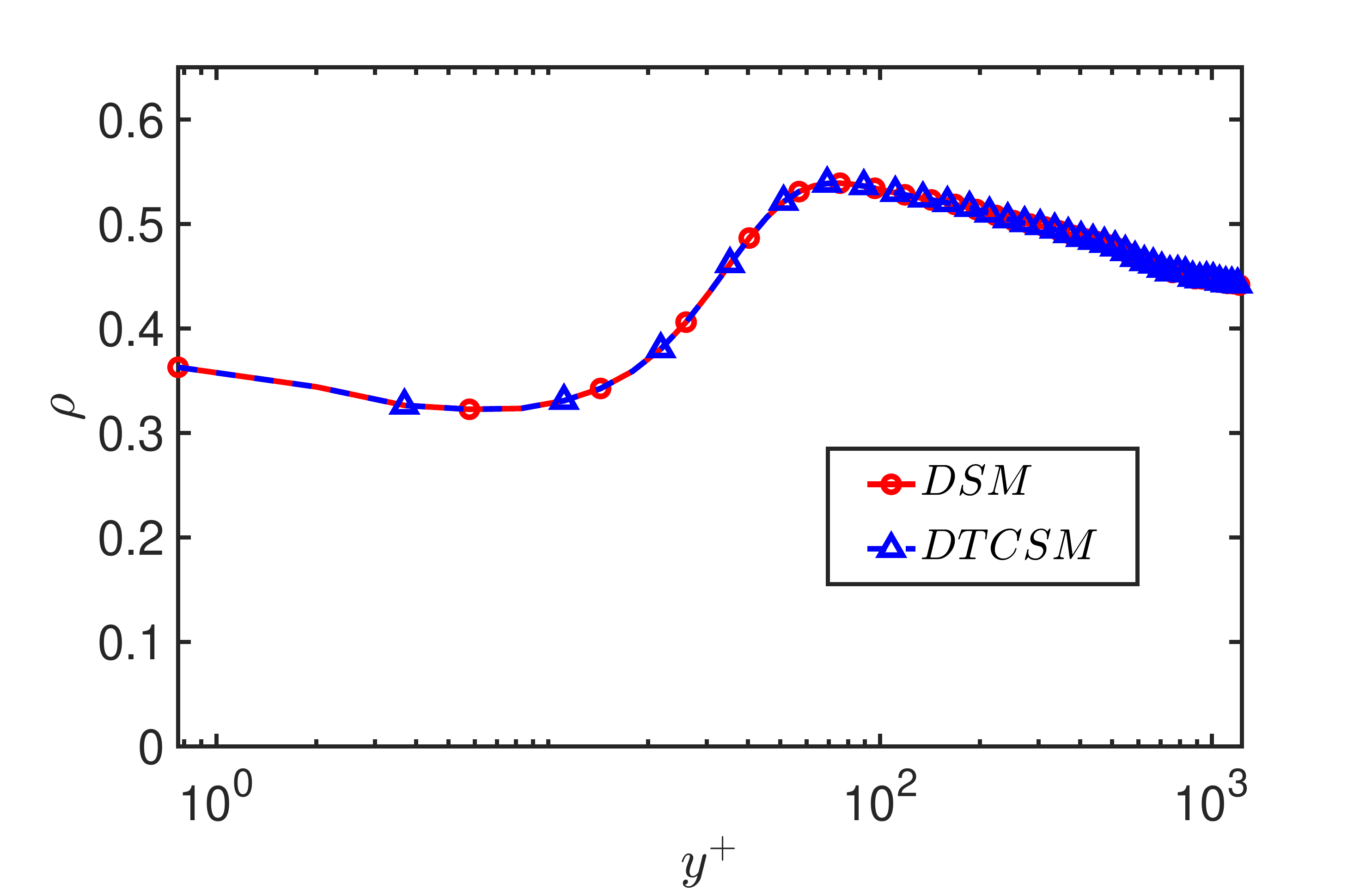}}
    \caption{ \textcolor{black}{ Dissipation rate correlations of DSM and DTCSM with respect to filtered DNS of turbulent channel flow at (a) $Re_{\tau} = 395$ and (b) $Re_{\tau}=2000$ respectively. Note that every third value has been marked for the higher Reynolds number case for visual clarity. }}
    \label{fig:aprioriscalar}
\end{figure}

At the \emph{a priori} level, DTCSM has exactly the same dissipation rate correlation (see Figure \ref{fig:aprioriscalar}) as DSM across the entire channel height. This is indeed expected since the only terms in DTCSM that contribute to dissipation are Smagorinsky-type terms. Although not shown, varying the filter width from two times the grid size to up to four times the grid size has \textcolor{black}{shown} little sensitivity in the qualitative trends for both stress tensor and dissipation correlations. A verification of the dissipative aspects of these models in LES is presented next in the \emph{a posteriori} \textcolor{black}{analysis} section.

\section{Asymptotic behavior in laminar flows and near solid walls}

In laminar flows with sufficient resolution, the \textcolor{black}{resolved stresses} approaches zero and 
consequently, the computed dynamic coefficients will also vanish (which can be confirmed 
from inspection of Eq. (\ref{eqn:dyndtcsm})).   

Near a solid wall, in the viscous sublayer, universal scalings for all three components of velocities have been well established \cite{kim1987turbulence}. Germano et al. \cite{germano1991dynamic} reported that the dynamic Smagorinsky model stresses in the 
near-wall region \textcolor{black}{followed the same behavior} with the distance from the wall. The eddy viscosity at the wall asymptotically reaches zero, and within the viscous sublayer it scales as $\nu_t \sim y^{+,3 }$ in this region, leading to the correct order of magnitude estimates for modeled subgrid stress in comparison to DNS. 

Assuming a test-filter kernel that does not operate in the wall-normal direction, and using velocity scalings from DNS as $u^{+} \sim y^{+}$, $v^{+} \sim y^{+,2}$ and $w^{+} \sim y^{+}$, the scalings for $L_{ij}$, $M_{ij}$ can be written in terms of distance from the wall in inner units. On solving the resulting system of equations in Appendix \ref{app:2}, it can be shown that for DTCSM, $C_{11} \sim y^{+,3}$, $C_{12} \sim y^{+,2}$, $C_{13} \sim 0 $, $C_{23} \sim y^{+,2}$. With these results, it can be easily inferred that the subgrid shear stress asymptotically reaches zero at the wall following the expected scaling $\tau^{sgs}_{12} \sim y^{+,3 }$. Thus, the behavior of the proposed model in near-wall region is in agreement with \textcolor{black}{the expected} asymptotic near-wall scalings. 

\section{A posteriori analysis in Canonical flows} 
Large eddy simulations with DTCSM are performed for decaying and forced homogeneous isotropic turbulence (HIT), and turbulent channel flow and the results are compared with DNS and LES with the DSM closure. \textcolor{black}{For these simulations, spatial averaging in homogeneous directions is performed to regularize the model coefficients for DSM and DTCSM.}

\subsection{Decaying homogeneous isotropic turbulence }

We perform DNS and LES of decaying HIT \citep{comte1971simple} at $\textit{Re}_{\lambda} = u_{rms} \lambda/\nu  = 70$. The initial turbulent field follows the spectrum in Passot and Pouquet \cite{passot1987numerical}, with $u_{rms} = 1$. For a triply-periodic box of size $(2 \pi )^3$, the DNS is performed on a grid of $128^3$ resolution while LES is performed using a coarser grid containing $32^3$ points. For HIT, the turbulent kinetic energy ($tke$) and  dissipation rate ($\epsilon$) are defined as
\begin{equation}
     tke = \langle \frac{1}{2} u_i u_i \rangle  \hspace{1mm} \mathrm{and} \hspace{1mm} \epsilon = \langle 2 \nu \overline{S}_{ij} \overline{S}_{ij} - \tau^{sgs}_{ij} \overline{S}_{ij} \rangle 
\end{equation}
\begin{figure}
     \centering 
     \includegraphics[width=0.65\textwidth]{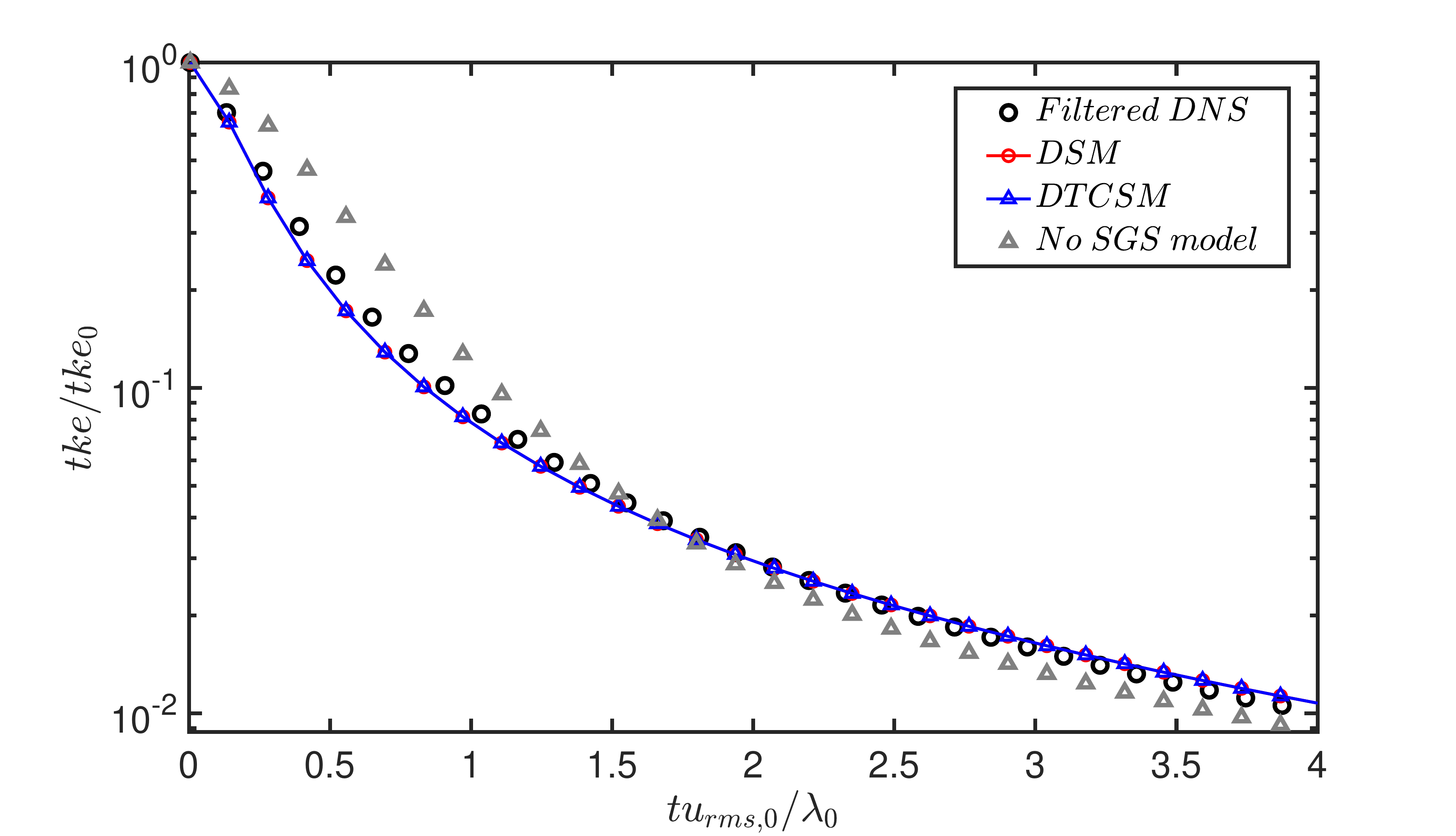}
         \caption{ \textcolor{black}{ Evolution of turbulent kinetic energy for decaying isotropic turbulence. }}
         \label{fig:hitenergy}
\end{figure}
\begin{figure}[!ht]
     \centering         \includegraphics[width=0.65\textwidth]{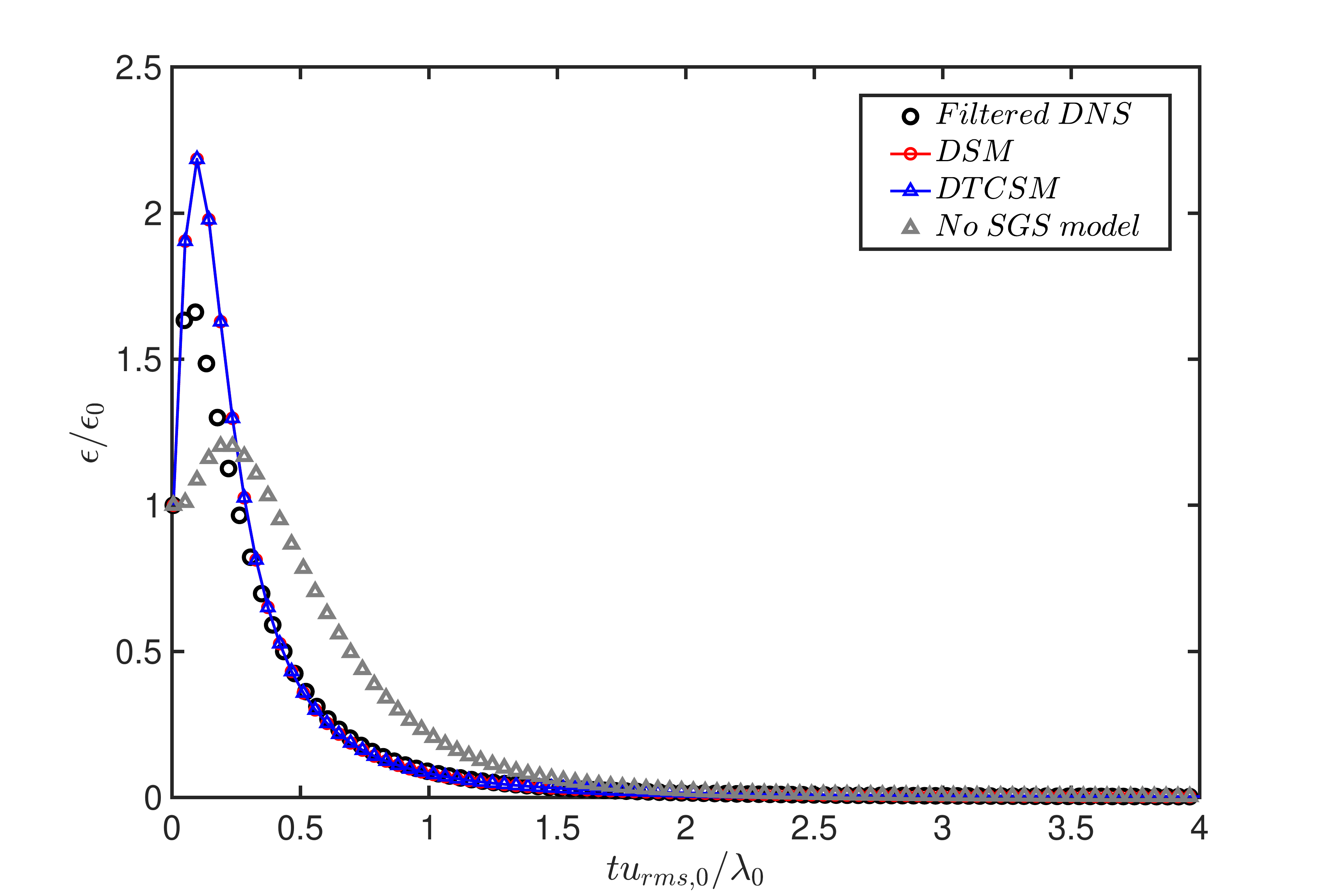}
         \caption{  \textcolor{black}{Evolution of the turbulent kinetic energy dissipation rate for decaying isotropic turbulence}. }
         \label{fig:hitdissipation}
\end{figure}
where $\langle \cdot \rangle$ is the volumetric-averaging operator. \textcolor{black}{Note that $tke_0$ and $\epsilon_0$ are the initial turbulent kinetic energy and its dissipation rate respectively.} In Figures \ref{fig:hitenergy} and \ref{fig:hitdissipation}, we compare the evolution of kinetic energy and the dissipation rate in LES to the filtered DNS. Since the calculation without an SGS model does not dissipate enough kinetic energy, it is apparent that the SGS models are needed to properly dissipate turbulent kinetic energy. The time evolution of both $tke$ and $\epsilon$ with both LES models are in excellent agreement with the filtered DNS after the initial transient (of up to one eddy turn-over time, due to a random-phase based initialization of the velocity field from the chosen energy spectrum). \textcolor{black}{The exponent of the decay of turbulent kinetic energy, $\alpha$, when $tke \sim t^{\alpha}$, is $\alpha \approx -1.35$, for both DSM and DTCSM, and is in reasonable agreement with that of the filtered DNS, $\alpha \approx -1.37 \;$ (filtered using a box filter of filter width equal to the LES grid size).} 
It is then apparent that DTCSM dissipates energy as well as DSM without any \emph{ad hoc} modifications to the 
dynamic procedure. It should be noted that some existing constant coefficient non-Boussinesq SGS models \citep{bardina1980improved,clark_ferziger_reynolds_1979} under-dissipate energy in such calculations and require augmentation from a Smagorsinky-model-type term.


\subsection{Forced homogeneous isotropic turbulence}
We now compare LES of forced HIT at $\textit{Re}_{\lambda} = 315$ with $128^3$ grid points with the filtered DNS of Cardesa et al. \cite{cardesa2017turbulent} performed with $1024^3$ grid points. A linear momentum forcing is applied to maintain constant turbulent kinetic energy in the system \citep{bassenne2016constant}. 

%

\begin{figure}[!ht]
    \centering
    \includegraphics[width=0.7\textwidth]{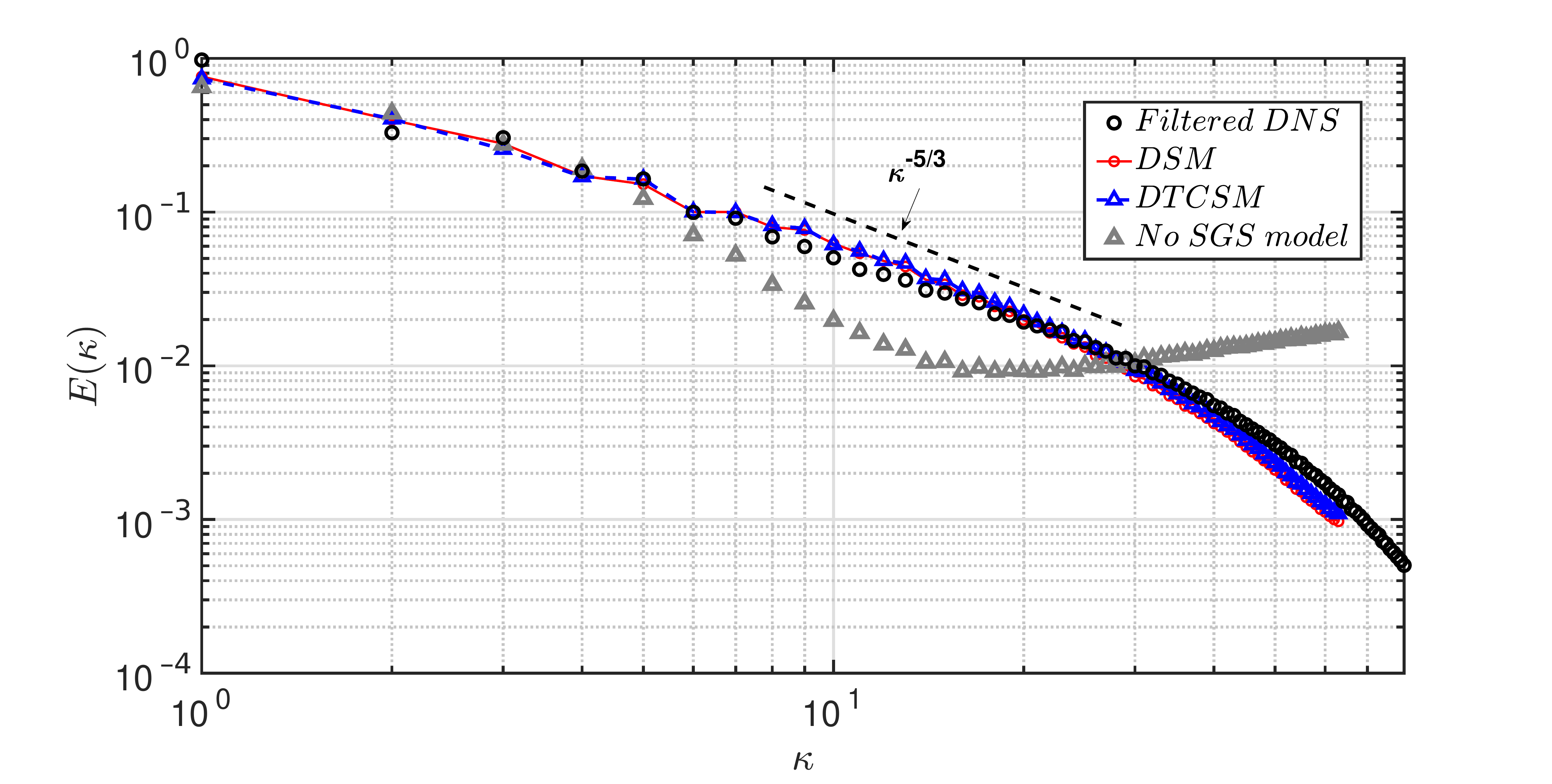}
    \caption{  Comparison of kinetic energy spectra, $E(\kappa)$, in forced isotropic turbulence between filtered-DNS and SGS models. The resolution of the filtered-DNS is at $1024^3$ compared to the LES resolution of $128^3$.  }
    \label{fig:KEforcedspec}
\end{figure}

Three-dimensional energy spectra are compared in Figure \ref{fig:KEforcedspec}, where the $\kappa^{-5/3}$ scaling is well recovered by both DSM and DTCSM. The \textcolor{black}{LES energy spectra} compare very well with the filtered DNS (evaluated using a box filter of filter width equal to LES grid size). This agreement suggests that the proposed model transfers energy from the largest scales to the inertial subrange as expected and hence on average do not show any scale-to-scale spurious energy transfer. Since the flow does not have large-scale global anisotropy, it is also expected that the performance of DSM would be similar to that of DTCSM, which is consistent with our results. 
\subsection{Homogeneous isotropic turbulence in the limit $\textit{Re}_{\lambda} \rightarrow \infty $}

\begin{figure}[!ht]
    \centering
    \includegraphics[width=0.7\textwidth]{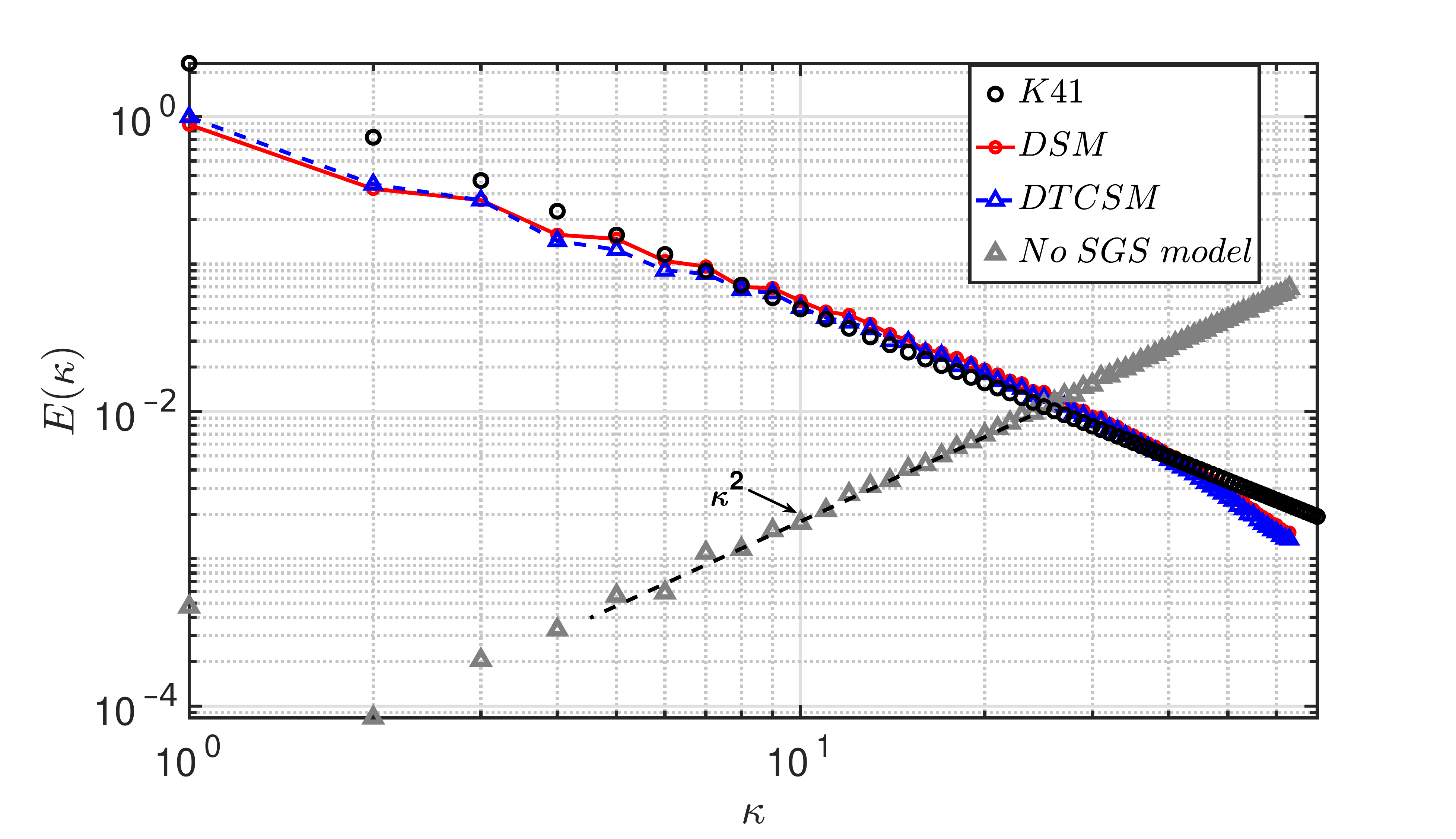}
    \caption{  Comparison of kinetic energy spectra, $E(\kappa)$, between filtered-DNS and SGS models as $Re_{\lambda} \rightarrow \infty $.  The LES resolution considered here is $128^3$. }
    \label{fig:KEinfspec}
\end{figure}
In the limit of $\textit{Re}_{\lambda} \rightarrow \infty$, multiple decades of the $K41$ scaling (i.e., $E \sim \kappa^{-5/3}$) are expected across the spectrum. From Figure \ref{fig:KEinfspec}, it is evident that the $K41$ scaling is well recovered with DTCSM (as well as DSM), further confirming both their dissipative and inter-scale energy transfer properties. 
The  $\kappa^2$ scaling in the absence of an SGS model is observed, consistent with the principle of equipartition of energy.

\subsection{Wall-modeled LES of channel flow }

In this section, we investigate the performance of DTCSM in a channel flow at $\textit{Re}_{\tau} = u_{\tau} \delta / \nu  = 4200$, where $u_{\tau}$ is the friction velocity and $\delta $ is channel half-height. Lozano-Dur{\'a}n and Jim{\'e}nez \cite{lozano2014effect} performed DNS of this flow with grids as refined as $\Delta x^+ = 12.8, \; \Delta y_{min}^+ = 0.31$ and $\Delta z^+ = 6.4$ in inner units \textcolor{black}{(approximately 10 billion grid points)}. For practical WMLES calculations, the grid resolutions are specified in outer units, and grid resolutions of $20-60$ points ($\Delta y^+ \sim 60 - 200$) across the boundary layer have been previously used \citep{lozano2019error}. In this work, we use 20 points per half-height of the channel. 


\begin{figure}[!ht]
    \centering
    \includegraphics[width=0.7\textwidth]{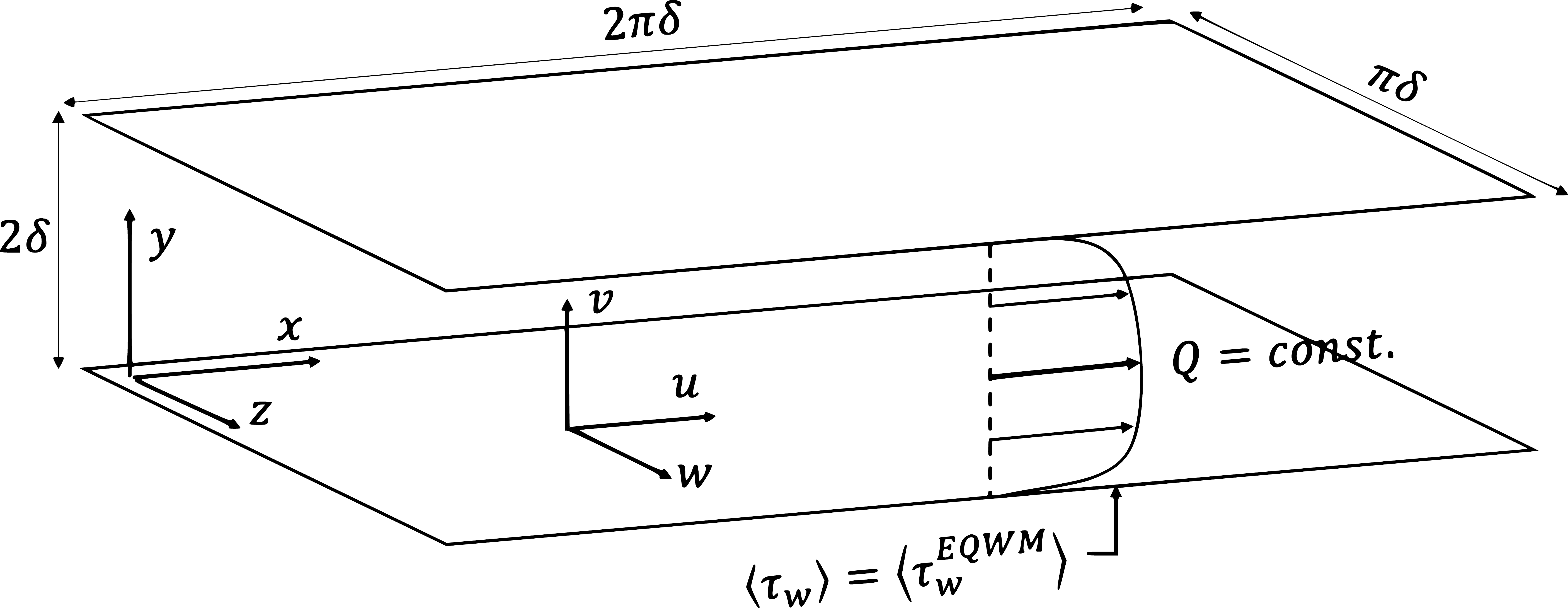}
    \caption{  Schematic of the channel-flow WMLES setup. }
    \label{fig:channelschematic}
\end{figure}

\begin{table}
\begin{tabular}{ p{2cm}p{0.8cm}p{0.8cm}p{0.8cm}p{0.8cm}p{1.0cm}p{1.2cm}p{1.0cm}  }
Simulation & $N_x$ & $N_y$ &$N_z$ & $\Delta x^+$ & $\Delta  y_{min}^+$ & $\Delta  y_{center}^+$ & $\Delta  z^+$\\
\hline
DNS   & 3072 & 1081 & 3072 & 12.8 & 0.31 & 10.7 & 6.4\\
WMLES & 128 & 40 & 64 & 200& 200 & 200 & 200\\
\end{tabular}
\caption{Simulation parameters for turbulent channel flow at $\textit{Re}_{\tau} = 4200$}
\label{table:reswmles}
\end{table}

Figure \ref{fig:channelschematic} shows the schematic of the channel flow and domain sizes in the three directions. Table \ref{table:reswmles} summarizes the simulation parameters used in the DNS reference and the present WMLES. \textcolor{black}{Unlike the DNS, the grid spacing is uniform and nearly isotropic in all three directions.} The channel is driven at constant mass flow rate to match the mass flow rate from the DNS.
{\color{black} The viscously dominated inner layer is not resolved, but rather, is represented through a wall model, and its effect imposed onto the outer LES by a wall shear stress boundary condition. The wall model for the channel flow simulations is the equilibrium wall model of Cabot and Moin \cite{cabot2000approximate}, which solves the steady thin boundary layer equations as a one-dimensional boundary value problem, given a matching location and wall-parallel velocity from the outer LES solution, in order to predict the wall shear stress. The matching location is chosen to be the second off-wall grid point (the distance of the matching location from the wall is henceforth  referred to as $y = h_{wm}$) in order to improve the accuracy of the wall shear stress prediction in finite-different, Cartesian-staggered meshes \cite{kawai2013dynamic}. However, as will be noted in Section \ref{sec:bump}, for unstructured Voronoi-HCP grids, matching at the first off-wall location was found to be sufficiently accurate. }



\begin{figure}[!ht]
    \centering
  \includegraphics[width=0.73\textwidth]{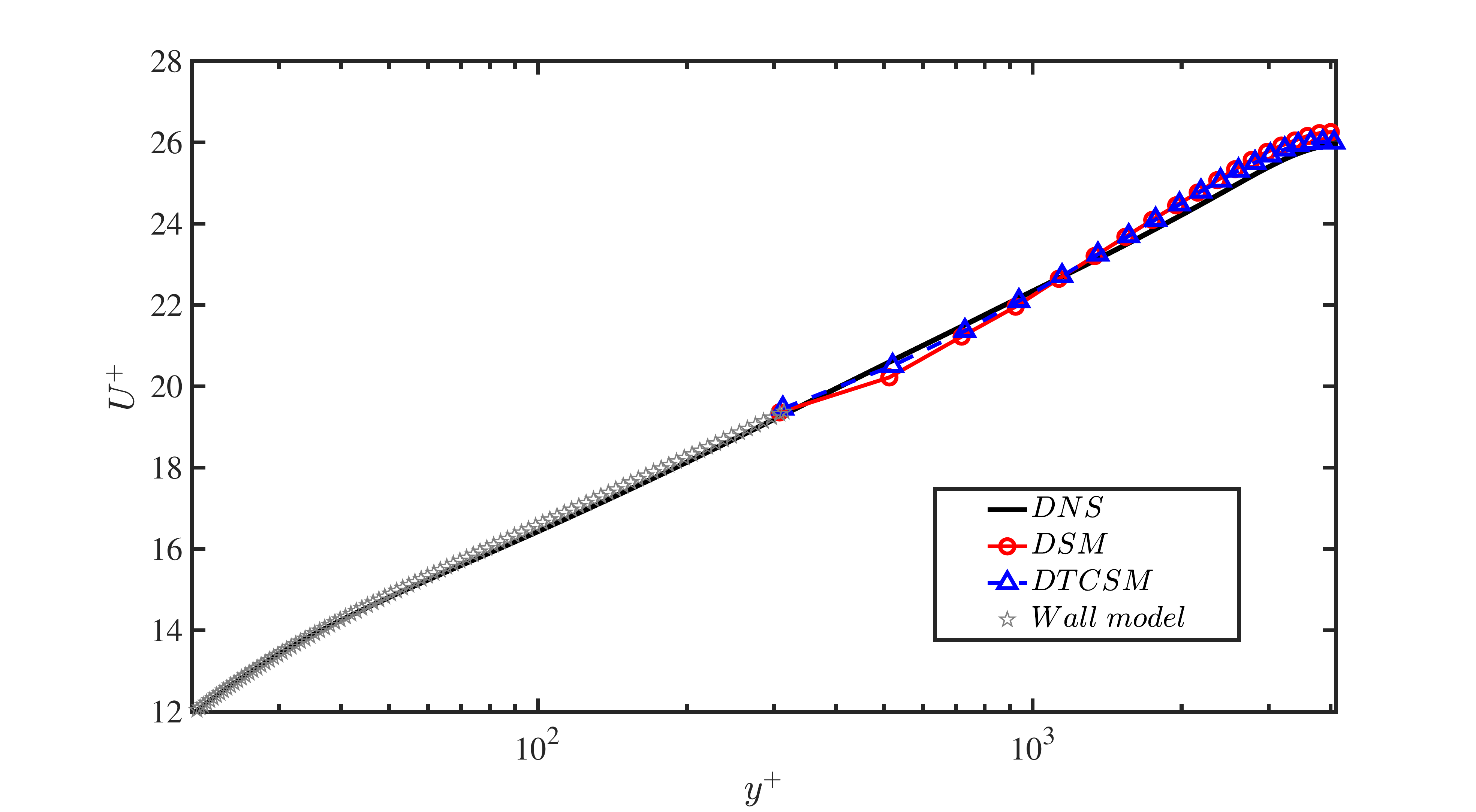} 
    \caption{ \textcolor{black}{ Wall-normal profiles of mean streamwise velocity for DSM and DTCSM in turbulent channel flow at $Re_{\tau} = 4200$. The mean profile predicted by the equilibrium wall model is plotted for wall normal distances less than the height of the matching location ($y \le h_{wm}$).} }
    
    \label{fig:meanu}
\end{figure}

\textcolor{black}{Kawai and Larsson \cite{kawai2012wall} argued that the energy-containing scales of the near-wall turbulence are, by definition, under-resolved and potentially inaccurate in the first grid point ($y^+ \sim 100$ in this work) on a WMLES grid. The observed behavior in this work with the chosen numerical scheme is that both DSM and DTCSM (which rely on extracting information from the resolved turbulent eddies) under-predict the subgrid-scale stresses, leading to larger velocity fluctuations which amplify the resolved Reynolds shear stress and flatten the slope of the mean velocity profile. Lozano-Dur{\'a}n and Bae \cite{lozano2019error} have previously shown convergence of the integrated errors upon grid-refinement in the mean velocity profile with respect to DNS for locations above the matching location. Thus, following Kawai and Larsson \cite{kawai2012wall}, Yang et al. \cite{yang2017llm}, only the mean velocity profile predicted by WMLES for $y \ge h_{wm}$ is plotted. For $y \le h_{wm}$, the mean profile predicted by the equilibrium wall model is considered more accurate and hence plotted. }
From Figure \ref{fig:meanu}, DTCSM is slightly better than DSM at predicting the mean streamwise velocity profile throughout the logarithmic region. Specifically, in the log layer, the error in the prediction of the K\'{a}rm\'{a}n constant ($\kappa = 0.38$ predicted by DNS) is \textcolor{black}{slightly lowered from $7\%$ error with DSM to $5\%$ with DTCSM}.  

\begin{figure}[!ht]
    \includegraphics[width=0.75\textwidth]{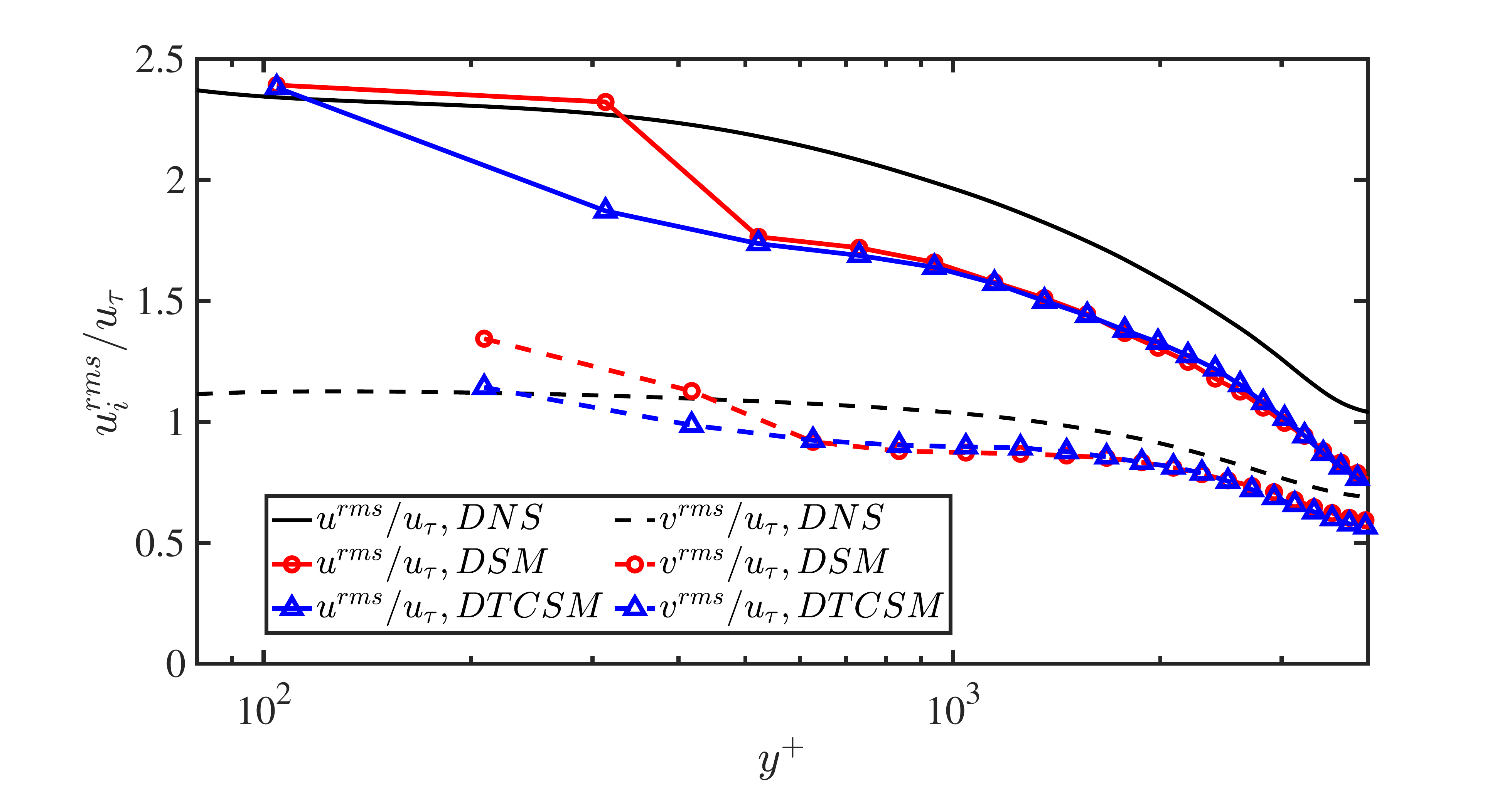}
    \caption{  Wall-normal profiles of turbulent intensities for DSM and DTCSM in turbulent channel flow, $Re_{\tau} = 4200$. The curves without symbols represent the mean intensities from DNS.  }
    \label{fig:intensities}
 \end{figure}

The over-prediction of the mean streamwise component of intensity in the near-wall region with respect to unfiltered DNS in LES of channel flows has been a concern; LES is expected to under-predict the intensity in order to be consistent with filtered DNS. In Figure \ref{fig:intensities}, it is observed that the streamwise and wall-normal components of turbulent intensities with DTCSM are lower than those with the DSM near the wall, which is a qualitative indicator that the prediction is improving when compared to turbulence fluctuations based on unfiltered DNS (explicitly filtered DNS data is not used here). It must be noted that improvements in the over-prediction of streamwise intensities have also been previously reported using a slip (Robin) boundary condition \textcolor{black}{in} a lower Reynolds number \textcolor{black}{channel flow} \cite{bae2018turbulence}. 

\begin{figure}[!ht]
    \centering
    \includegraphics[width=0.7\textwidth]{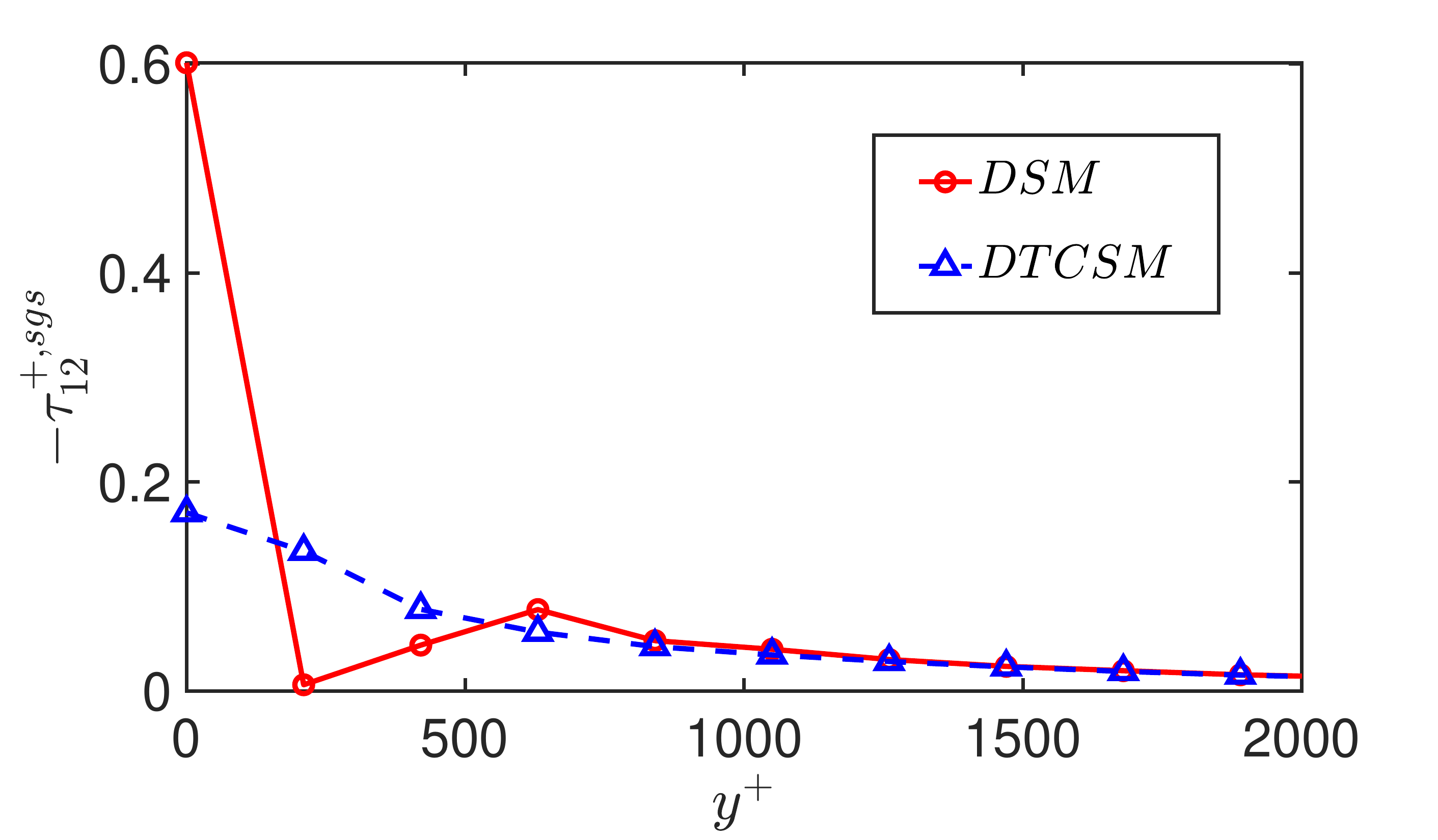}
    \caption{  Wall-normal profiles of mean SGS shear stress for DSM and DTCSM in WMLES of turbulent channel flow at $Re_{\tau} = 4200$. Note that a \textcolor{black}{Neumann boundary condition ($d \tau^{sgs} /dy  = 0$ ) is used for the SGS stresses at the wall.}  }
    \label{fig:sgswmles }
\end{figure}

For a turbulent channel flow driven at $\textit{Re}_{\tau}$ as low as 395, Morinishi and Vasilyev \cite{morinishi2002vector} observed that DSM requires {ad hoc} clipping in the near-wall region. A similar trend is observed at $\textit{Re}_{\tau} = 4200 $ in our simulations \textcolor{black}{using DSM} (Figure \ref{fig:sgswmles }), where the subgrid stresses in the second off-wall point are clipped. No such clipping was necessary with the DTCSM for this flow. \textcolor{black}{It should be noted that at the wall, a Neumann boundary condition is used for the SGS stresses, which is motivated by the fact that in coarse wall-modeled calculations, the resolved turbulent stresses are formally non-zero \cite{bose2014dynamic}. The Neumann boundary condition was hence used on the subgrid-scale shear stresses to account for the non-zero turbulent stresses in the equilibrium wall model. } 

\section{Wall-modeled LES of flow over Gaussian bump}
\label{sec:bump}
\begin{figure}[!ht]
    \centering
    \includegraphics[width=0.7\textwidth]{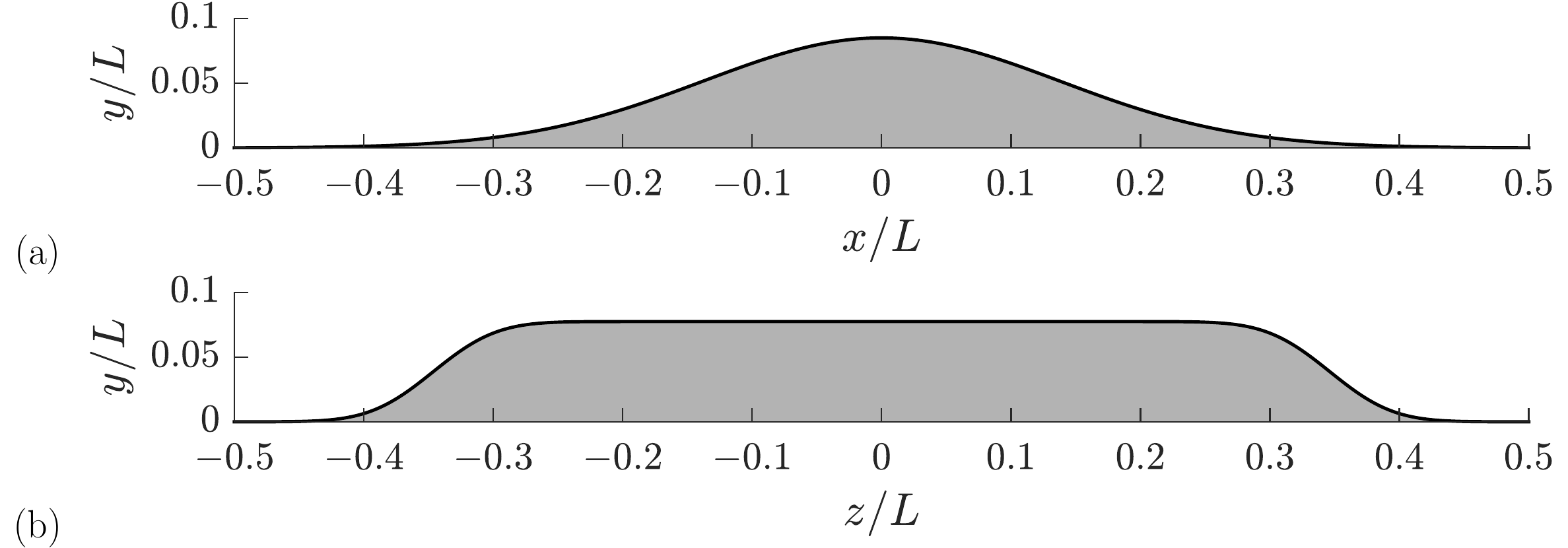}
    \caption{  Cross-sections of the three dimensional bump geometry. The geometry has side walls at $z/L = \pm 0.5$ and a top wall at $y/L = 0.5$. Note that the spanwise-periodic bump has the same cross-sectional profile as the three-dimensional bump, however, the spanwise direction is periodic with $z/L$ spanning from $0- 0.08$, and the top wall is at $y/L = 1.0$. }
    \label{fig:2dbumpfig}
\end{figure}

\begin{figure}[!ht]
    \centering
    (a)
    \includegraphics[width=0.9\textwidth]{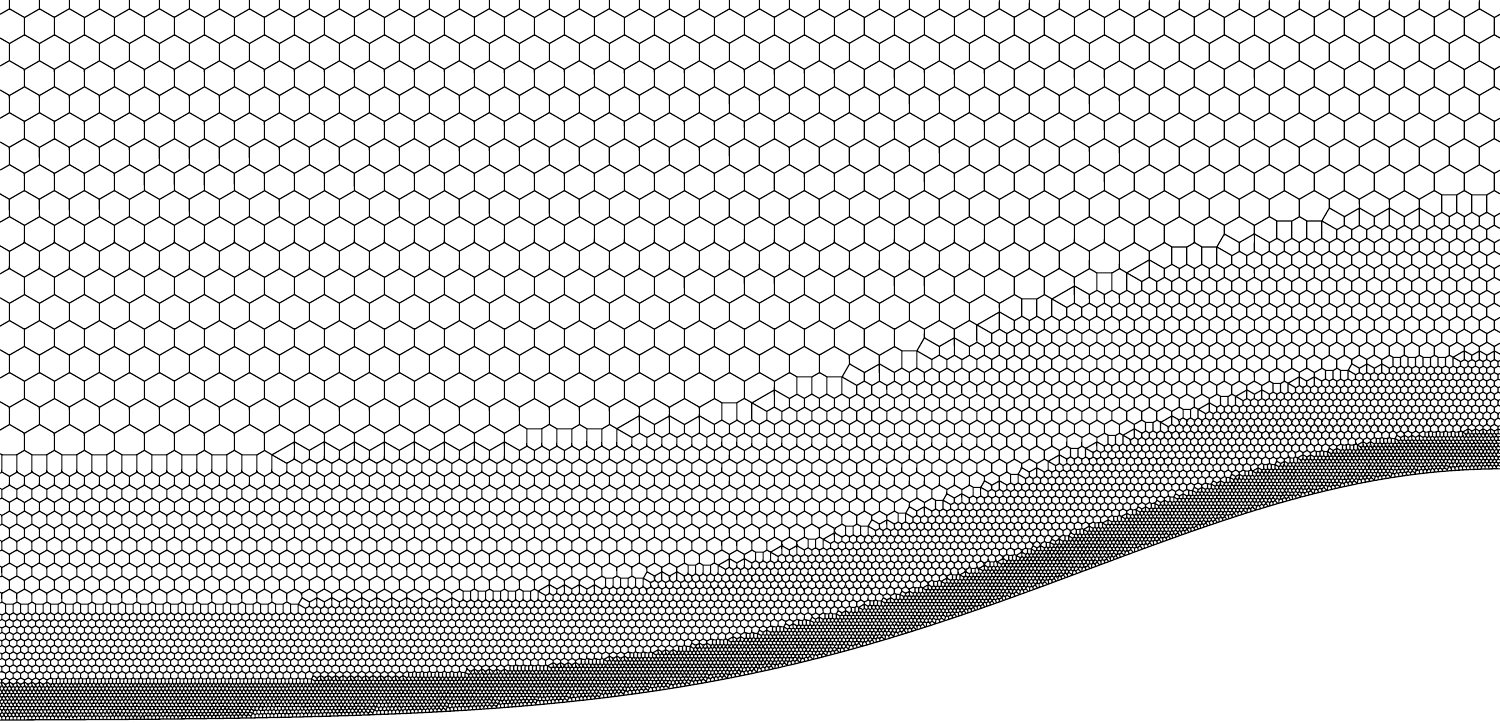}
    
    \vspace{0.3cm}
    
    (b)
    \includegraphics[width=0.9\textwidth]{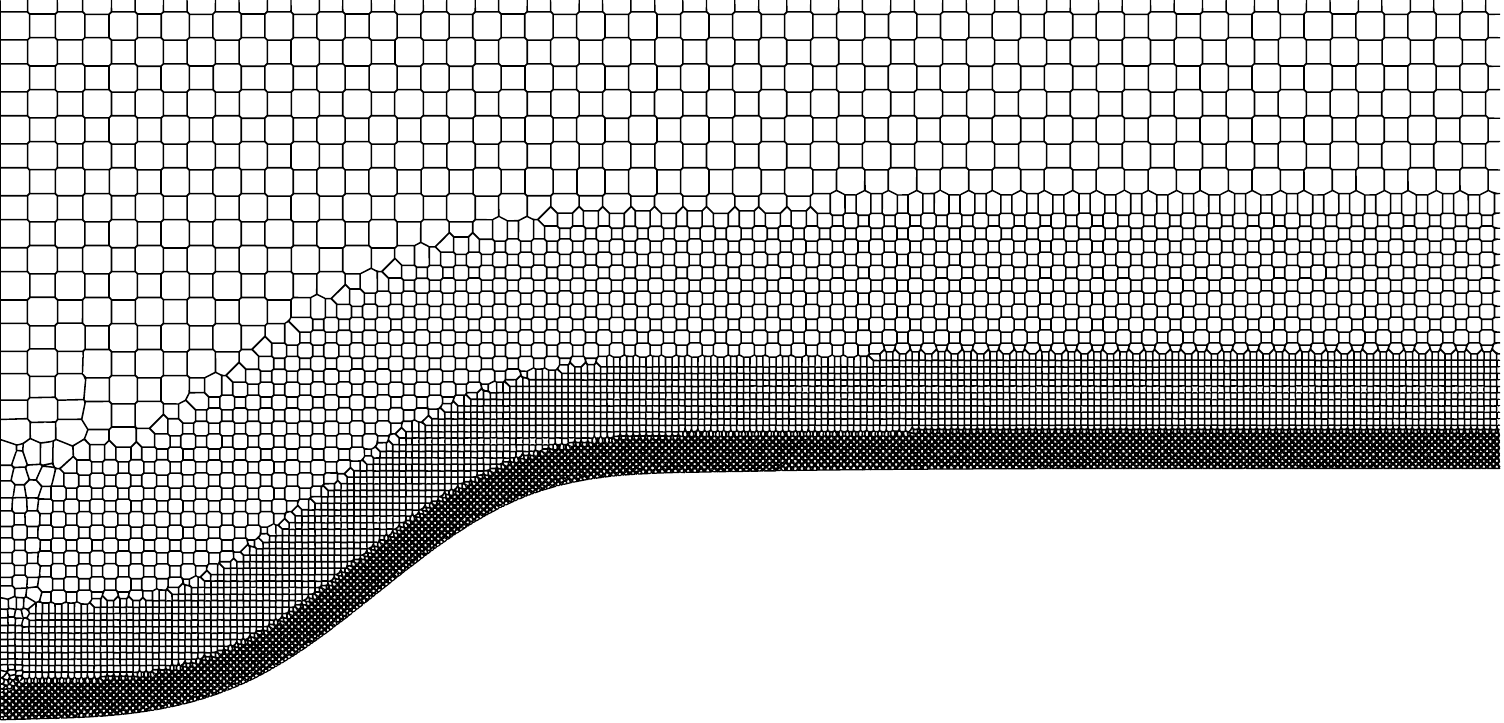}
    
    \caption{\color{black} Cross-sections of the coarse mesh for the experimental geometry (a) at the plane $z/L=0$, along the centerline of the bump, and (b) at the plane $x/L=0$, along the apex of the bump. The streamwise extent in (a) is $x/L\in [-0.5,0]$ showing the fore side of the bump, with the aft side being essentially symmetric. The spanwise extent in (b) is $z/L\in [-0.5,0]$ showing the left half of the domain while the right half is essentially symmetric. Both cross-sections show the lower half of the vertical extent of the domain. The top half of the domain is meshed uniformly up to the top wall. Three layers of isotropic refinement are visible adjacent to the bump surface. The isotropic refinement layers continue all the way to the inlet and outlet in either streamwise direction. Control volumes far from domain boundaries and refinement transitions consist of tessellated truncated octahedra.}
    \label{fig:3dbumpmesh}
\end{figure}

Next, the subgrid-scale models are deployed in the simulation of flow over a wall-mounted Gaussian bump (also known as the Boeing speed-bump \cite{williams2020experimental}). The geometry of the bump was proposed by Boeing and Williams et al. \cite{williams2020experimental}. The bump surface is defined by an analytical expression, $h(x,z)$, written as
\begin{equation}
    h(x,z) = \frac{h_0}{2} \, e^{-(x/x_0)^2} \,\left\{ 1+ {\rm erf}\left[\left(\frac{L}{2}-2 z_0 - |z| \right) / z_0 \right] \right\} ~,
    \label{eqn:bump}
\end{equation}
where $x$ and $z$ are the streamwise and spanwise coordinates, respectively. The bump width, $L$, is used to scale the other dimensions of the bump, as well as define a Reynolds number, $Re_L$. Here, $h_0 = 0.085L$ is the maximum height of the bump, and $x_0 = 0.195L$ controls the Gaussian decay of the surface in the streamwise direction. \textcolor{black}{Figures \ref{fig:2dbumpfig} and \ref{fig:3dbumpmesh} are schematics of the bump geometry and computational mesh as a function of streamwise and spanwise coordinates. }

In this flow, the turbulent boundary layer is subjected to both favorable and adverse pressure-gradients leading to the formation of a separation bubble. Experimental results \cite{gray2022experimental} have revealed approximate Reynolds number independence for pressure and skin-friction coefficients. Due to the presence of side walls and the associated mean three-dimensionality in the flow, and the availability of experimental data, the Boeing speed bump is a good candidate for assessing the predictive capability of LES models in a smooth-body separation.

Unlike many widely studied canonical separated flows, such as backward-facing steps or bumps which have geometrically-imposed separation points \citep{balakumar2015dns,cho2021wall,seifert2002active,driver1985features}, the present geometry features a smooth-body separation whose location and extent is significantly more difficult to predict computationally. In fact, several recent investigations have observed that some existing wall and subgrid-scale models struggle to correctly predict the occurrence/location of separation \cite{iyer2021wall,whitmorebump}. 

Uzun and Malik \cite{uzun2021high} performed a quasi-DNS of the spanwise-periodic variant geometry. For the spanwise-periodic geometry, the height of the bump surface is given by the simplified equation $h(x) = h_0\,{\rm exp}(-x^2/x^2_0)$. In the separated region, the surface pressure is found to agree well with the experiment of Williams et al. \cite{williams2020experimental} along the centerline $z=0$. Further, the skin friction measurements at mid-span for the three dimensional configuration are in good agreement with the spanwise-periodic case \citep{gray2022experimental}. Uzun and Malik \cite{uzun2021high} refer to their simulation as a quasi-DNS since it obeys the resolution requirements of DNS in the near-wall region and most of the attached boundary layer; however, it is more comparable to LES resolution in the outer part of the boundary layer.

For the spanwise-periodic configuration, the inlet velocity profile, located at $x/L = -1.0$, is a mean profile sampled from a RANS computation \cite{uzun2021high}. 
The inflow boundary layer is steady before undergoing a numerical transition to turbulence; thus, there is a development length in the WMLES simulation over the region $-1.0 < x/L < -0.8$. Free-stream conditions are set at the top boundary for this case. On the other hand, the three dimensional bump configuration has a plug flow inlet at $x/L = - 1.0 $ with the side and top boundary conditions treated as inviscid walls to account for the wind tunnel walls. The outlet is located at $x/L =2.5$ and $x/L = 1.5$ in the spanwise-periodic and experimental configurations respectively. A non-reflecting characteristic boundary condition with constant pressure is applied at the outlet. \textcolor{black}{For these simulations},  \textcolor{black}{ an algebraic formulation of the equilibrium wall-stress model, in which the assumed mean velocity profile is $C^1$ continuous is used. Details of the compressible formulation of the equilibrium wall model can be found in \cite{lehmkuhl2018large}. It should be noted that first-point matching has been used in these simulations (in our previous numerical experiments, no log-layer mismatch was observed with first point matching on Voronoi-HCP grids for this solver)}. \textcolor{black}{Finally, temporal averaging is performed to regularize the model coefficients in DSM and DTCSM for these calculations.}


For this flow, the quantities of interest are the skin friction coefficient ($C_f$) and pressure coefficient ($C_p$) which are defined as,
\begin{equation}
    C_f = \frac{\tau_{w}}{ 1/2 \rho_{\infty} U^2_{\infty}} \hspace{5pt} \mathrm{and} \hspace{5pt}  C_p = \frac{ p - p_{\rm ref}}{ 1/2 \rho_{\infty} U^2_{\infty}}.
\end{equation}
where $U_{\infty}$, $\tau_{w}$, $p$ and $p_{\rm ref}$ are the inlet free-stream velocity, mean wall-stress, wall pressure, and reference pressure, respectively. The reference pressure is taken to be the pressure at the wall at $x/L=-0.83$ to match the reference pressure used by Williams et al. \cite{williams2020experimental}.
    
Both the spanwise-periodic and three-dimensional bumps have the same background mesh spacing, $\Delta/L = 0.01$. The grid is a centroidal Voronoi diagram generated from a hexagonally close-packed lattice of seed points. While refining the grid, the control volumes (CVs) are refined homothetically by factors of 2 in layers near the wall boundaries. Each successive mesh has CVs twice as fine \textcolor{black}{in each direction} as those of the previous mesh. This refinement approach injects a thin layer (10 cells thick) of twice refined cells in all directions into the regions immediately adjacent to the bump. Lloyd iterations are employed to smooth the mesh at transitions in resolution; the result is a nearly centroidal mesh. The same refinement strategy for the spanwise-periodic and three-dimensional configurations facilitates comparison between them.
Three computational mesh \textcolor{black}{resolutions} are \textcolor{black}{considered} in the present work, details for which are provided in Tables \ref{table:resbump2} and \ref{table:resbump3}.

\begin{table}
\begin{tabular}{ p{1.5cm}p{1.5cm}p{1.5cm}p{1.5cm} }
Mesh & $N_{cv}$ & max $\Delta / L$  & min $\Delta / L$ \\
\hline\noalign{\vspace{3pt}}
Coarse  & $3$ Mil. & $0.01$ & $1.3 \times 10^{-3}$ \\
Medium  & $12$ Mil. & $0.01$ & $6.3 \times 10^{-4}$ \\
Fine    & $52$ Mil. & $0.01$ & $3.1 \times 10^{-4}$ \\
\end{tabular}
\caption{Mesh parameters for the spanwise-periodic bump case at $Re_L = 2\times 10^6$.}
\label{table:resbump2}
\end{table}

\begin{table}
\begin{tabular}{ p{1.5cm}p{1.5cm}p{1.5cm}p{1.5cm} }
Mesh & $N_{cv}$ & max $\Delta / L$  & min $\Delta / L$ \\
\hline\noalign{\vspace{3pt}}
Coarse  & $29$ Mil. & $0.01$ & $1.3 \times 10^{-3}$ \\
Medium  & $117$ Mil. & $0.01$ & $6.3 \times 10^{-4}$ \\
Fine    & $452$ Mil. & $0.01$ & $3.1 \times 10^{-4}$ \\
\end{tabular}
\caption{Mesh parameters for the three dimensional bump case at $Re_L = 3.41\times 10^6$.}
\label{table:resbump3}
\end{table}

\subsection{Spanwise-periodic bump at $Re_L = 2 \times 10^6 $ }

In this section we compare WMLES results with quasi-DNS data for the spanwise-periodic geometry at upstream Reynolds number, $Re_L = \rho_{\infty} U_{\infty} L / \mu_{\infty} = 2 \times 10^6 $  where $\mu_{\infty} \; \mathrm{and}\; \rho_{\infty} $ are the free-stream dynamic viscosity and density respectively. 

\begin{figure}[!ht]
    \centering
    \subfigure[]{
    \includegraphics[width=0.48\textwidth]{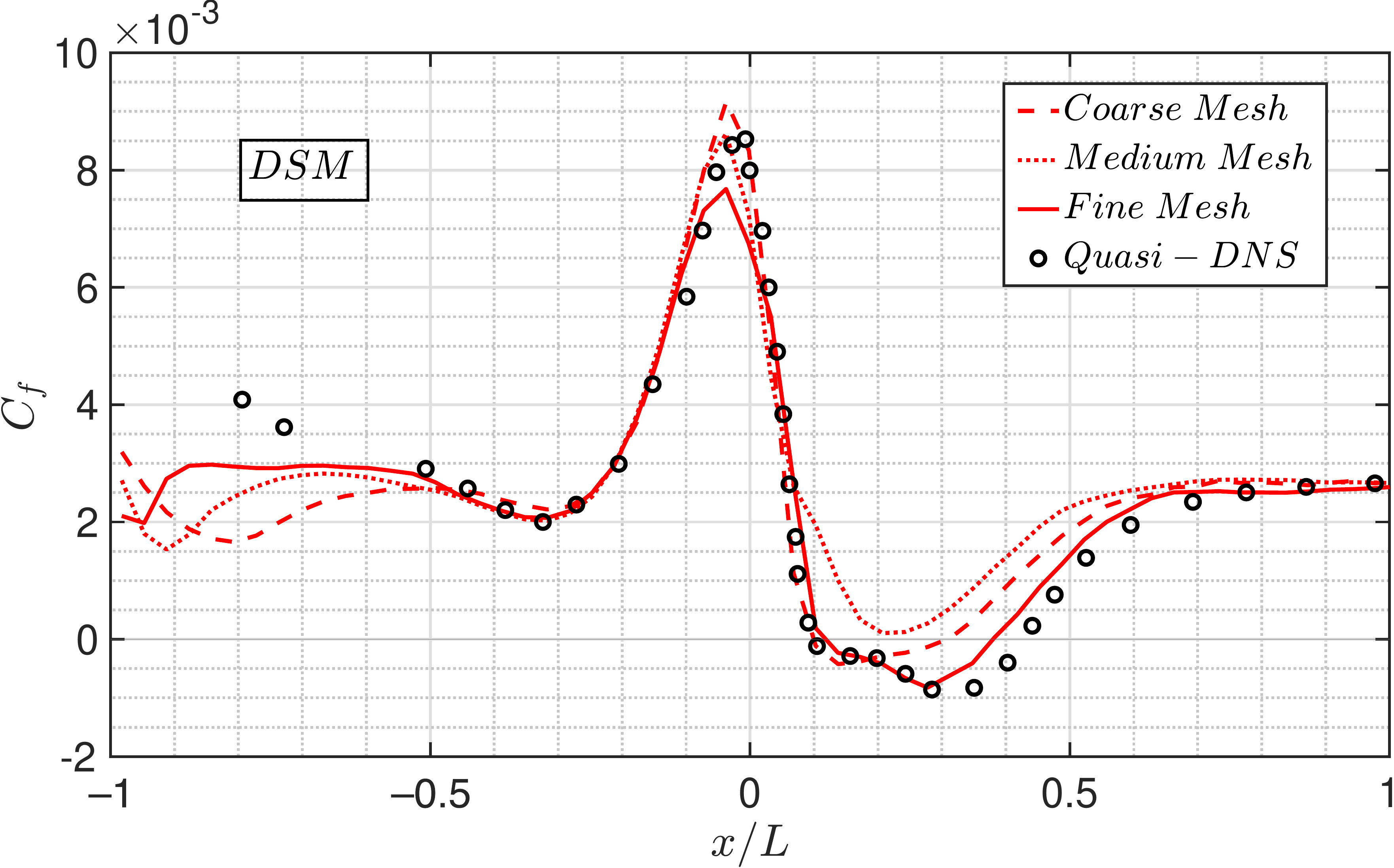}}
    \subfigure[]{
    \includegraphics[width=0.48\textwidth]{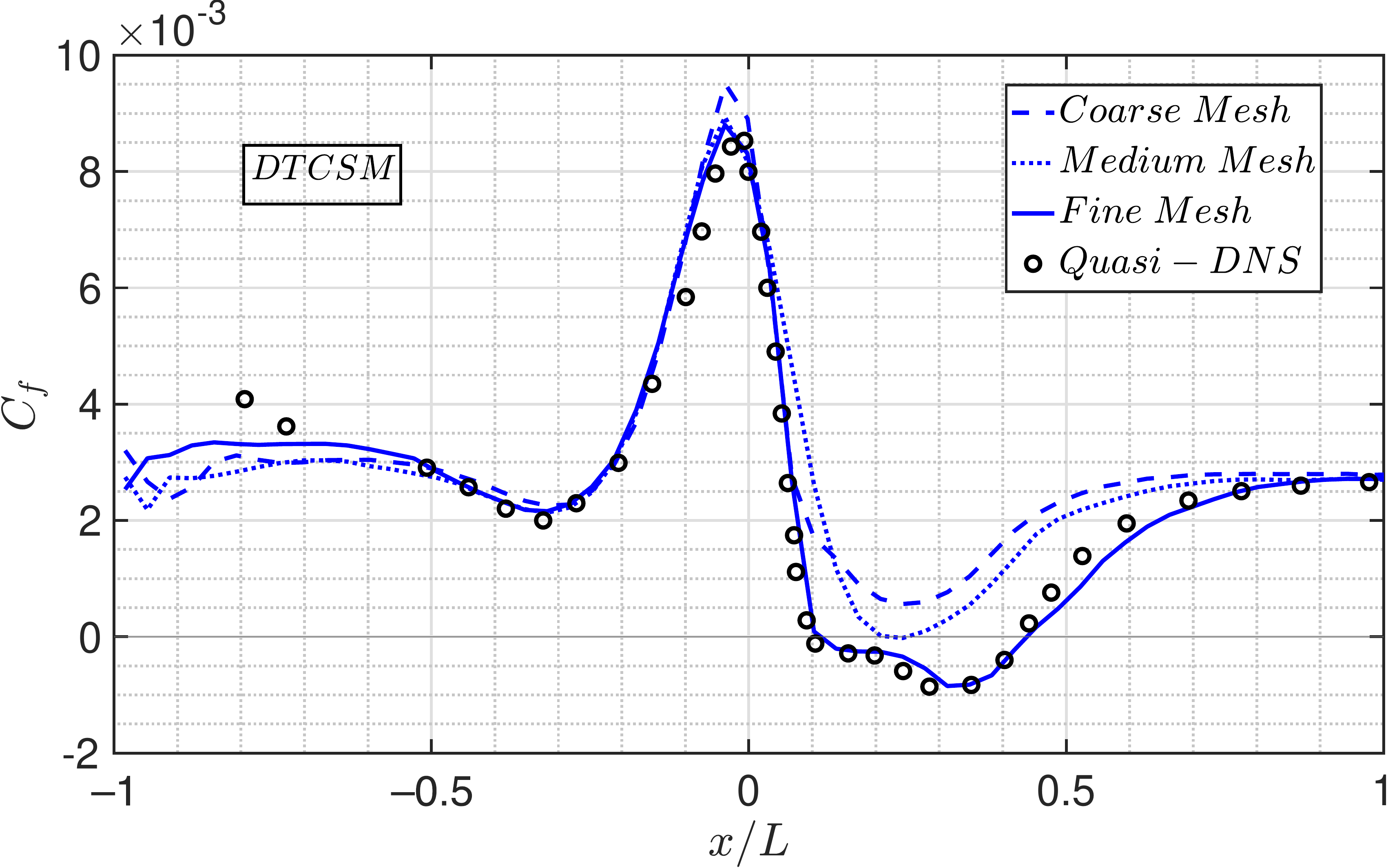}}
    \subfigure[]{
    \includegraphics[width=0.48\textwidth]{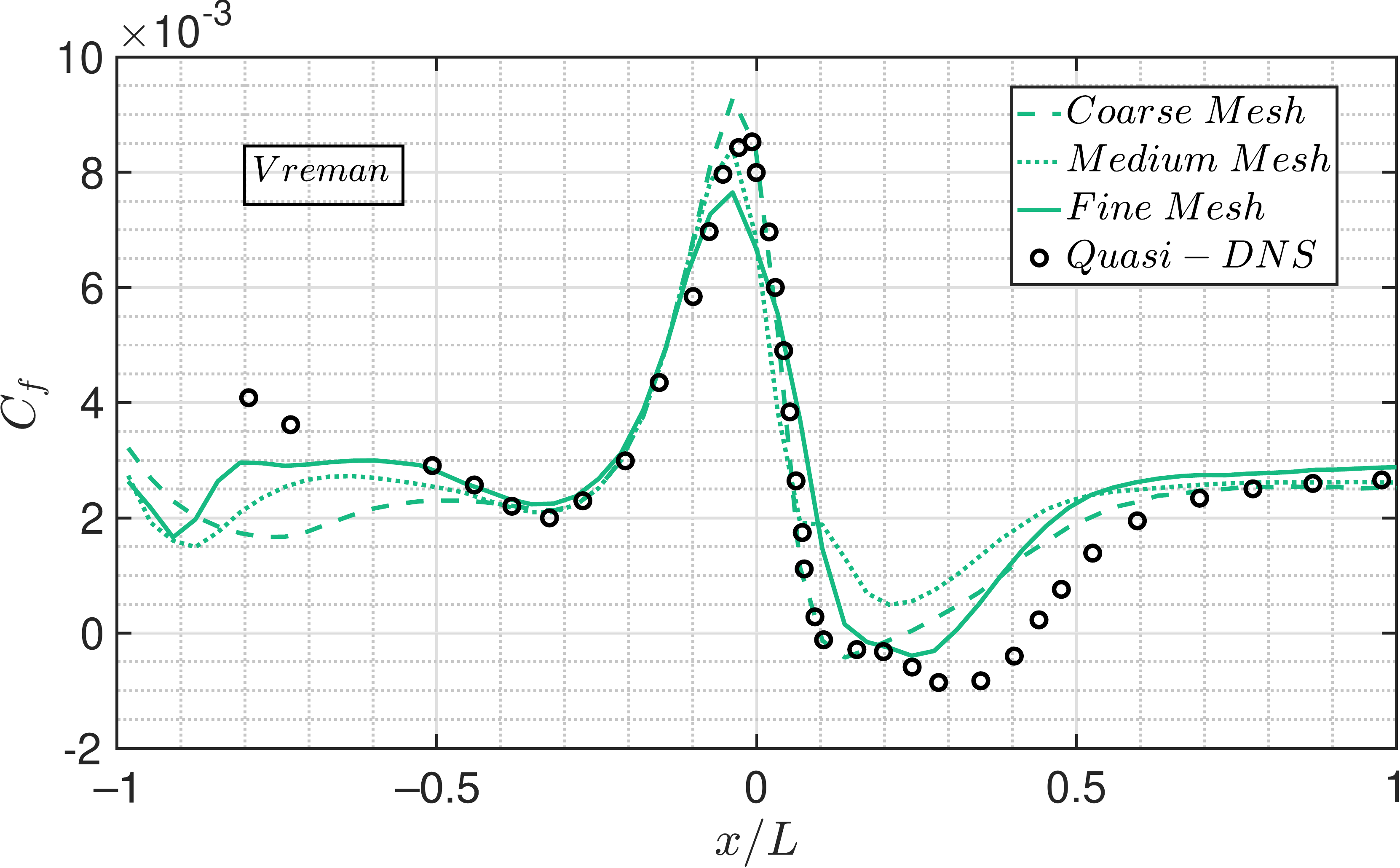}    }
    \subfigure[]{
    \includegraphics[width=0.48\textwidth]{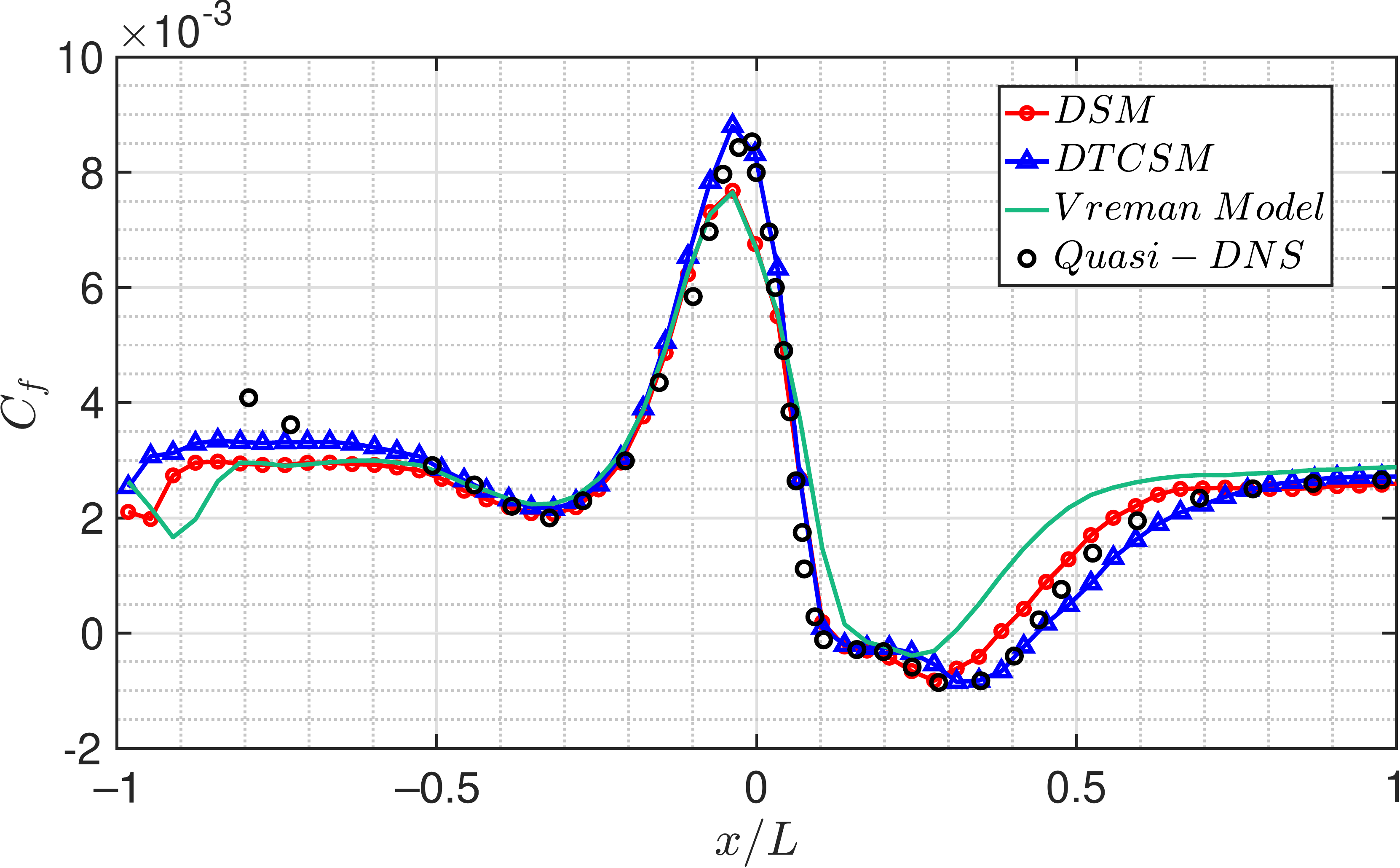}}
    \caption{  Streamwise distribution of the surface friction coefficient for the spanwise-periodic bump at $Re_L = 2 \times 10^6$ for (a) DSM, (b) DTCSM and (c) Vreman Model. The equilibrium wall model is applied and three mesh resolutions are shown. Finally, sub-figure (d) provides a comparison of the three SGS models for the fine mesh. The black symbols represent the quasi-DNS of Uzun and Malik \cite{uzun2021high}. }
    \label{fig:2dcf}
\end{figure}

In Figure \ref{fig:2dcf}, skin friction coefficients are compared across the three meshes for DSM, DTCSM and the 
Vreman model. 
The friction coefficient at the inlet is different from that of the quasi-DNS likely due to the difference in inlet boundary conditions (more details \textcolor{black}{can be found} in Whitmore et al. \cite{whitmorebump}). Similar to previous observations \cite{whitmorebump}, the coarse mesh results predict flow separation (albeit slightly under-predicted) using DSM. On refinement, the separation is diminished and then reappears on further refinement up to the fine mesh. At the fine mesh resolution, the separation bubble size is in good agreement with DNS for both DSM and DTCSM. It is noteworthy that DTCSM shows a monotonic convergence 
toward the DNS $C_f$, particularly in the region of separation. This is in contrast to the results from the DSM and Vreman 
models, which both predict the separation at coarse resolution, but show the medium resolution results moving away from the DNS solution. The non-monotonic convergence of WMLES towards DNS/experimental results has also been observed in the past with DSM and Vreman model in \textcolor{black}{simulations of the high-lift aircraft configuration} \cite{gocsubgrid,goc2021arb} where both models over-predicted the total lift in the linear region of the lift curve upon initial mesh refinement, and the predictions improved upon further refinement. 


Since this flow is driven by a favorable pressure gradient in the upstream region of the apex of the bump, there is a substantial 
increase in the skin friction. DSM under-predicts the peak of skin friction in this flow at the fine mesh resolution (when separation is correctly observed). However, DTCSM recovers the peak of the skin friction accurately while also \textcolor{black}{correctly} predicting the extent of separation. The peak of friction occurs in the region of maximum flow acceleration and anisotropic shear rates, which \textcolor{black}{apparently DTCSM is better equipped to account for, by incorporating the effect of anisotropy of large scale structures on the subgrid-scale stresses.}

\begin{figure}[!ht]
    \centering
     \subfigure[]{
     \includegraphics[width=0.48\textwidth]{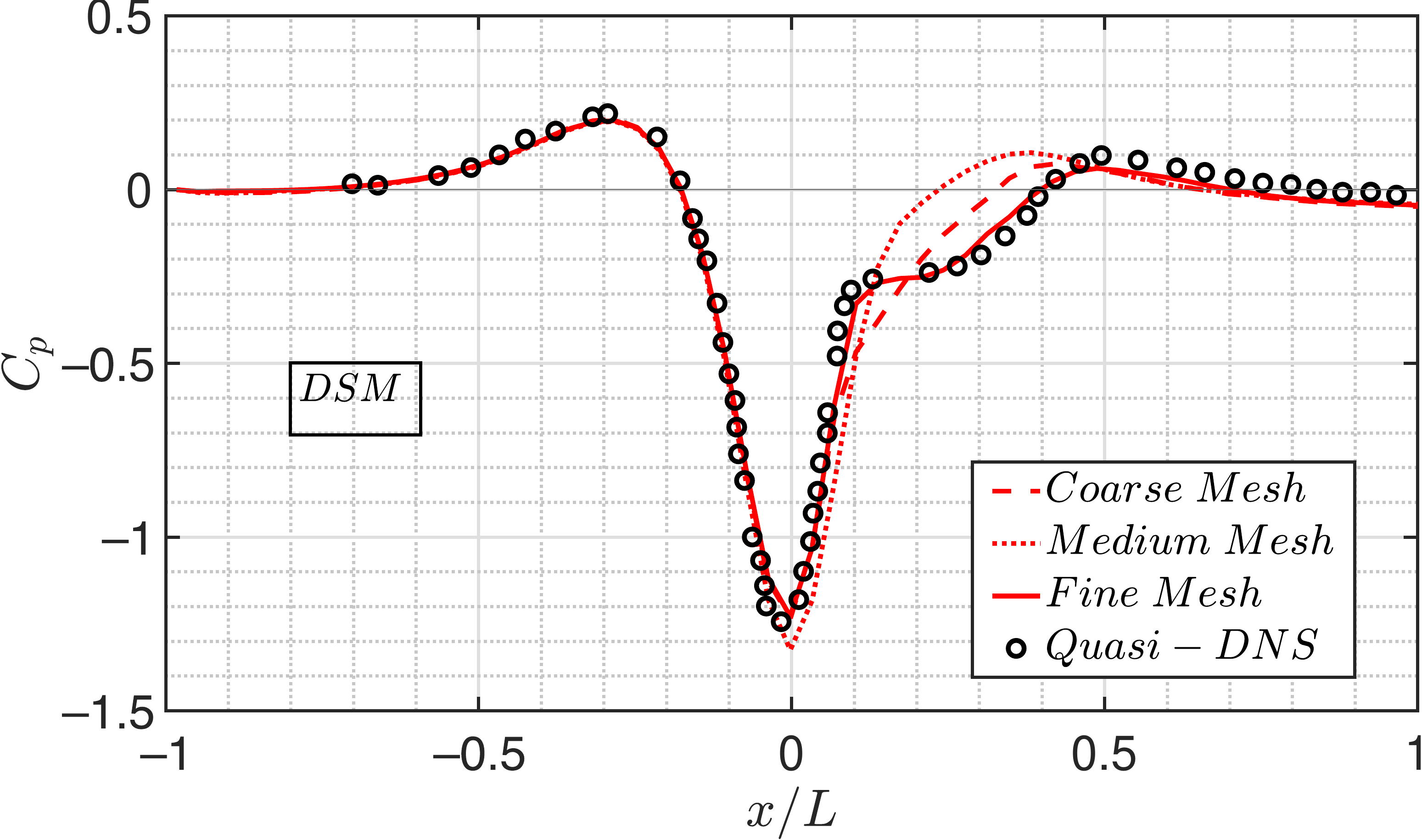}}
     \subfigure[]{
     \includegraphics[width=0.48\textwidth]{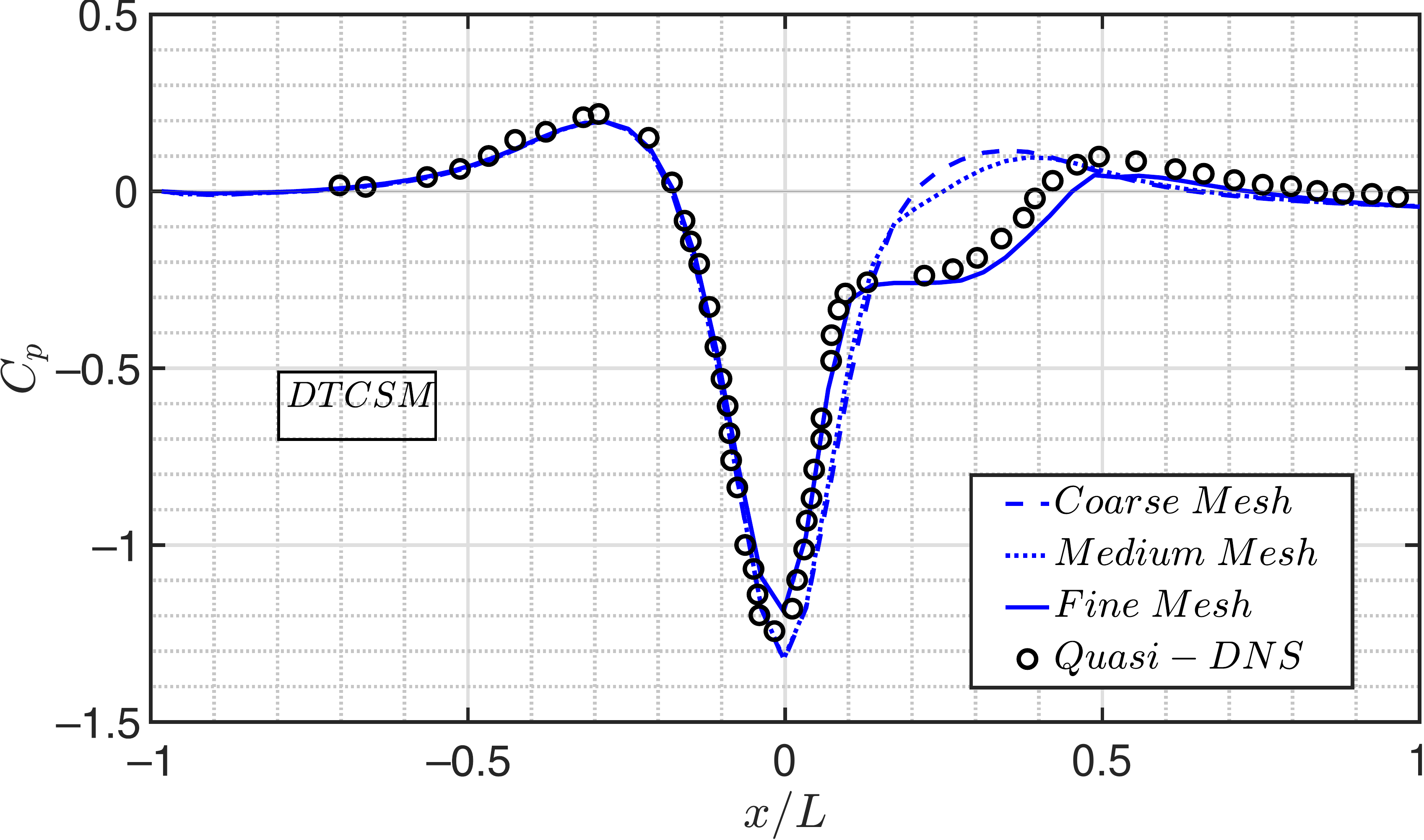}}
     \subfigure[]{
     \includegraphics[width=0.48\textwidth]{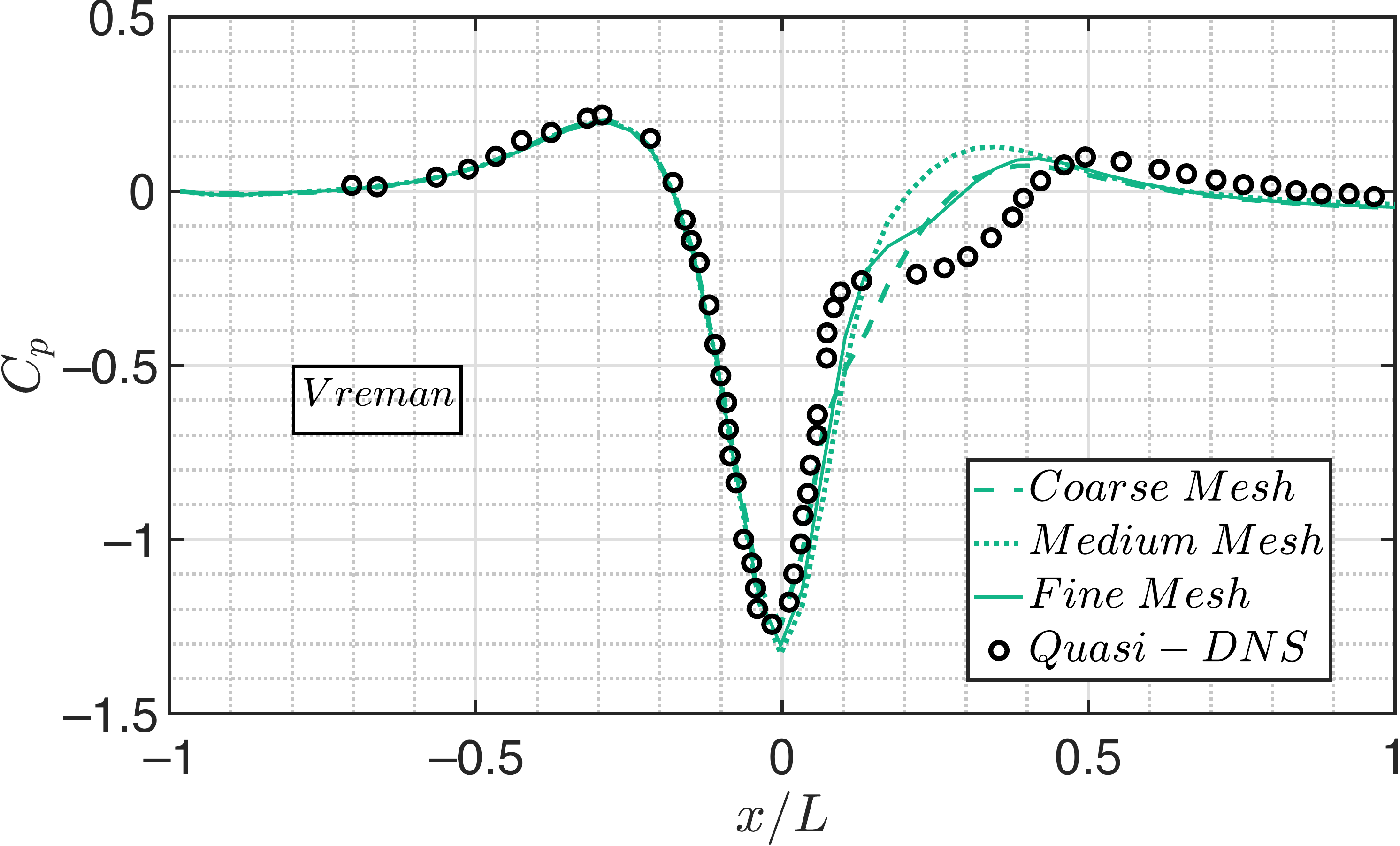}}
     \subfigure[]{
     \includegraphics[width=0.47\textwidth]{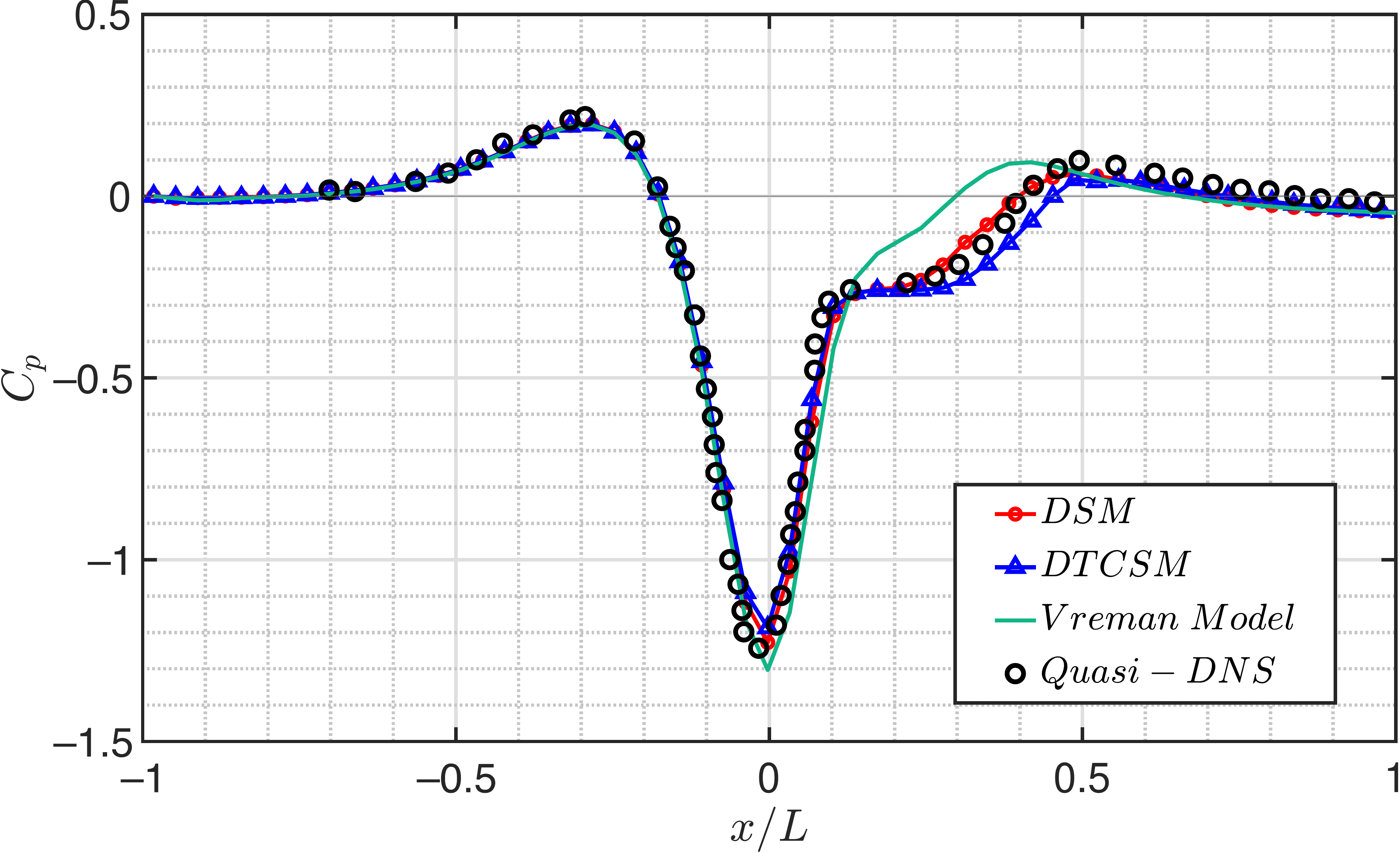}}
     \caption{  Streamwise distribution of the surface pressure coefficient for the spanwise-periodic bump at $Re_L = 2 \times 10^6$ for (a) DSM, (b) DTCSM and (c) Vreman Model. The equilibrium wall model is applied and three mesh resolutions are shown. Finally, sub-figure (d) provides a comparison of the three SGS models for the fine mesh. The black symbols represent the quasi-DNS of Uzun and Malik \cite{uzun2021high}. }
    \label{fig:2dcp}
\end{figure}

Figure \ref{fig:2dcp} compares the pressure coefficients across the models and mesh sizes.
The size of the separation bubble predicted by the Vreman model is smaller and compares less favorably than both dynamic models, even for the most refined grid. This is apparent from the very small flattened region of $C_p$ downstream of the bump apex in Figure \ref{fig:2dcp} for the Vreman model and also the region of negative $C_f$ in Figure \ref{fig:2dcf}. 

\subsection{Three-dimensional bump at $Re_L = 3.41 \times 10^6 $ }

WMLES results are presented for the full experimental geometry given in Eq. (\ref{eqn:bump}) at a higher Reynolds number of $Re_L = 3.41 \times 10^6$. Williams et al. \cite{williams2020experimental} and Gray et al. \cite{gray2022experimental} have performed experimental measurements of the surface pressure and skin friction respectively for the three-dimensional bump. Due to the tapering effect of the bump in the spanwise direction which relieves the pressure in the span, two counter-rotating vortices are formed in the separation region \cite{williams2020experimental}.
Iyer and Malik \cite{iyer2021wall} observed no separation in their WMLES calculations with the Vreman model with up to 450 Mil. CVs. Previous RANS efforts \cite{iyer2021wall,williams2020experimental} using the SA-Linear model and the SA-QCR correction model have also been unsuccessful in capturing the correct extent of the separation bubble (Iyer et al. \cite{iyer2021wall} observed weak separation using SA-QCR \textcolor{black}{RANS} model). The peak in the pressure in the spanwise direction at the apex of the bump was also absent in these studies. 

\begin{figure}[!ht]
     \centering
    \subfigure[]{
    \includegraphics[width=0.48\textwidth]{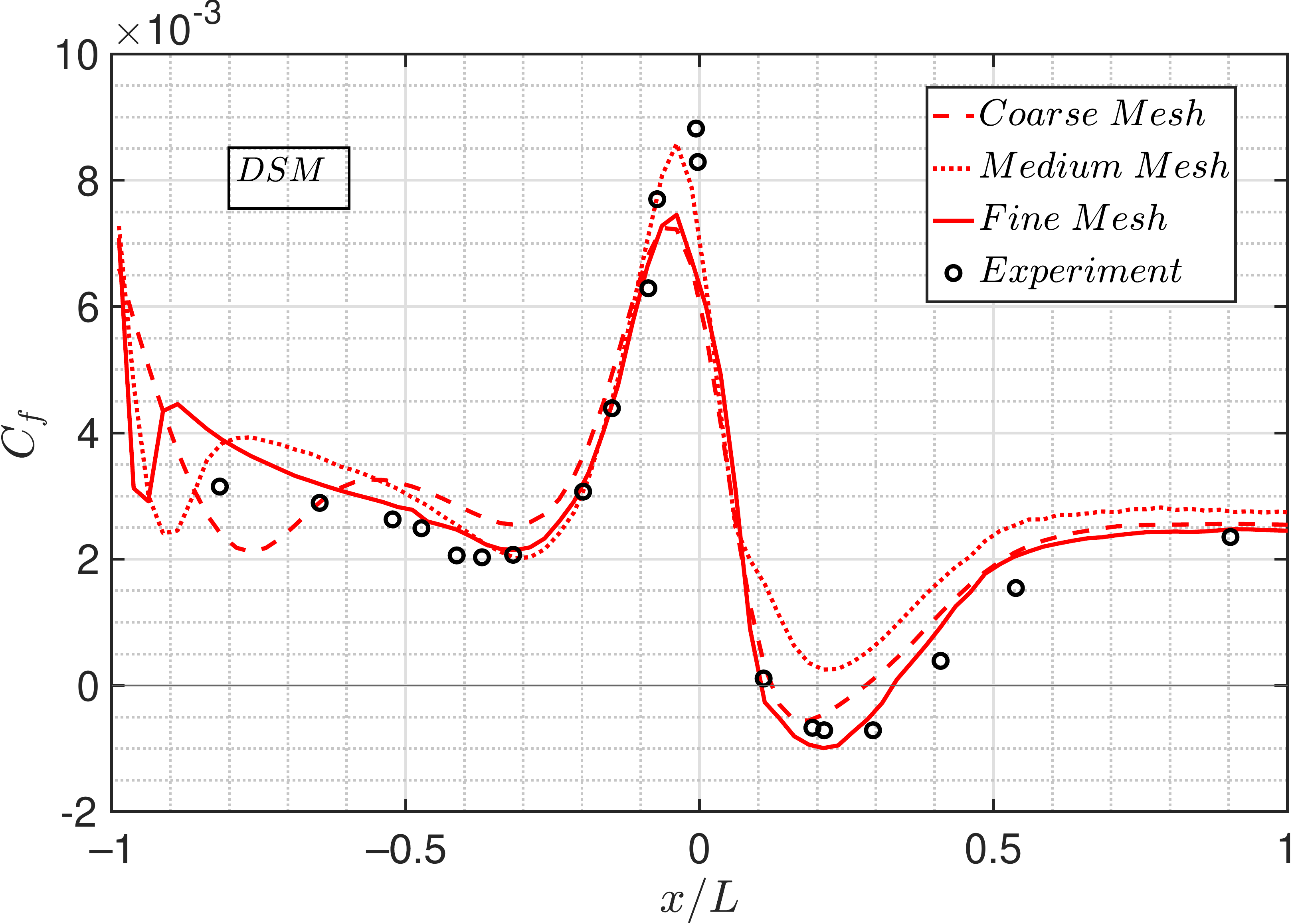}}
    \subfigure[]{
    \includegraphics[width=0.48\textwidth]{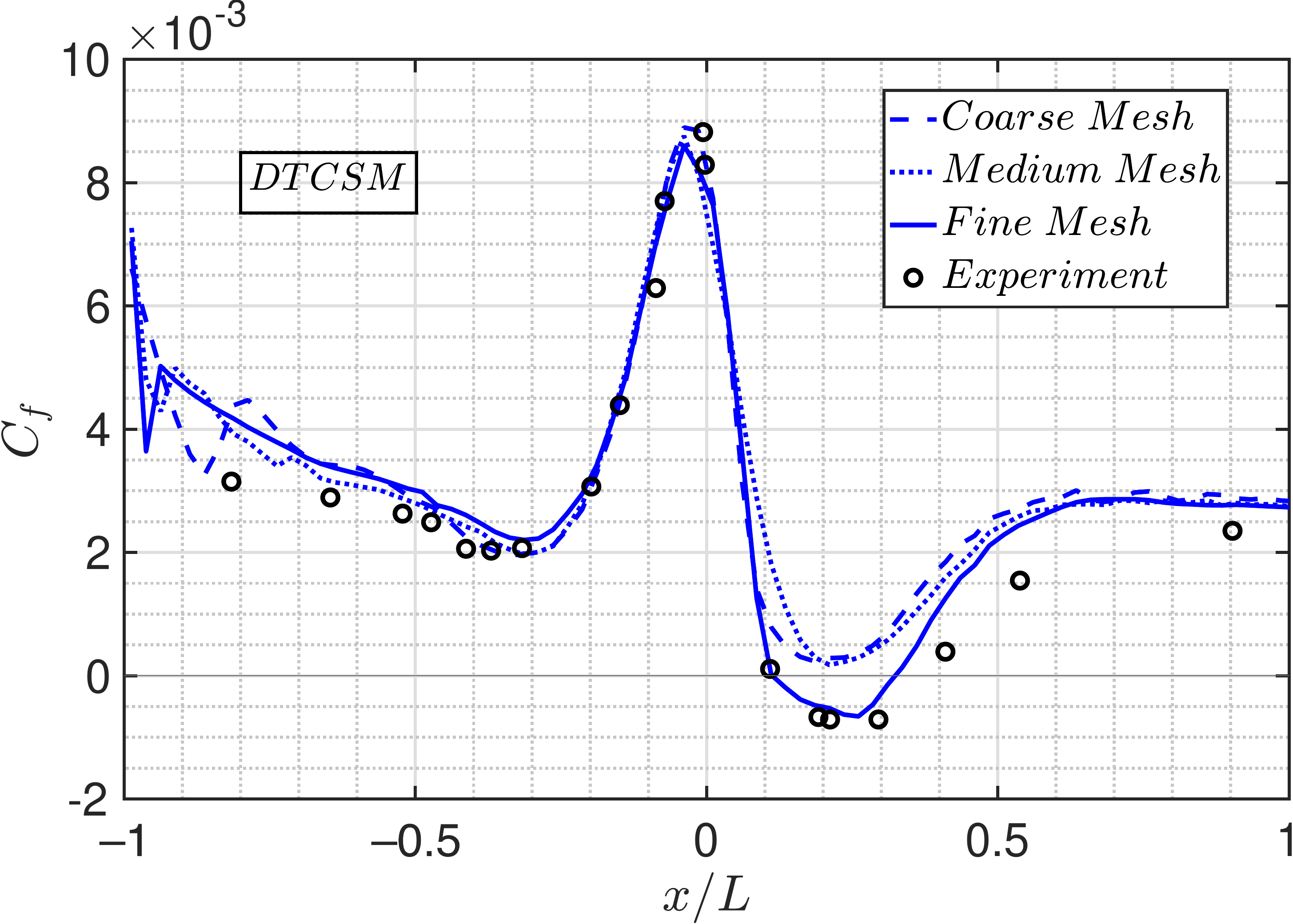}}
    \subfigure[]{
    \includegraphics[width=0.48\textwidth]{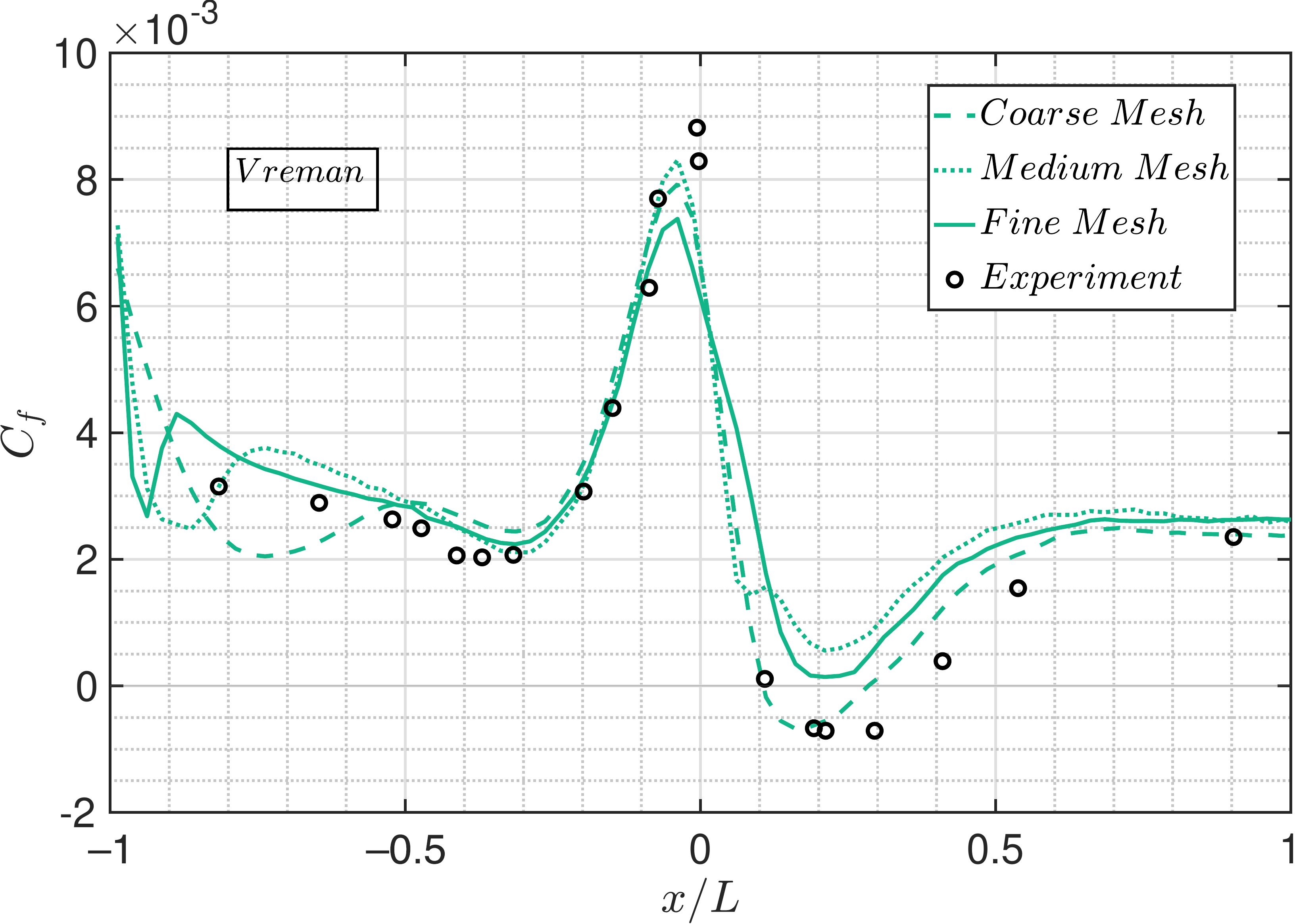}}
    \subfigure[]{
    \includegraphics[width=0.48\textwidth]{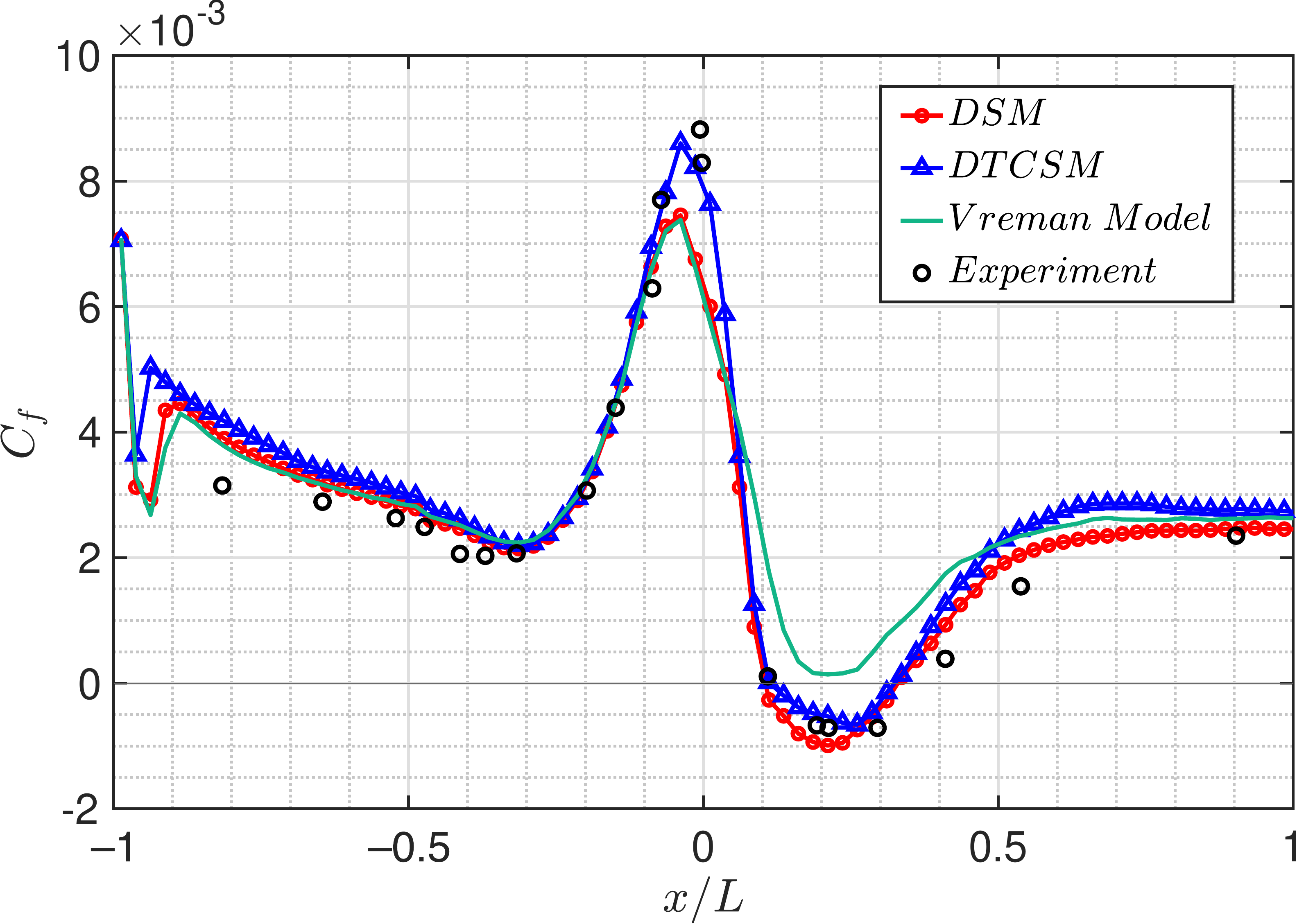}}
   \caption{  Streamwise distribution of the mid-span surface friction coefficient for the three dimensional bump at $Re_L = 3.41 \times 10^6$ for (a) DSM, (b) DTCSM and (c) Vreman model. The equilibrium wall model is applied and three mesh resolutions are shown. Finally, sub-plot (d) compares the prediction of $C_f$ by the three models on the fine mesh. The black dots represent the experiments performed by Gray et al. \cite{gray2022experimental}. }
    \label{fig:3dcf}
\end{figure}

As is the case with the spanwise-periodic bump, the skin friction coefficient profile suggests that DSM predicts flow separation (albeit smaller than experiment) on the coarsest grid unlike DTCSM (see 
Figure \ref{fig:3dcf}). On refinement, the separation bubble fails to appear for both DSM and DTCSM. After further refinement, the resolution of the nearest-to-wall cell center in the upstream region of the bump drops to about $y^+ \sim 30$. 
For this fine mesh, the separation bubble reappears for DSM and appears for the first time for DTCSM. The size of the separation bubble is in agreement with the measurements of Gray et al. \cite{gray2022experimental}. 

\begin{figure}[!ht]
    \centering
    \subfigure[]{
    \includegraphics[width=0.48\textwidth]{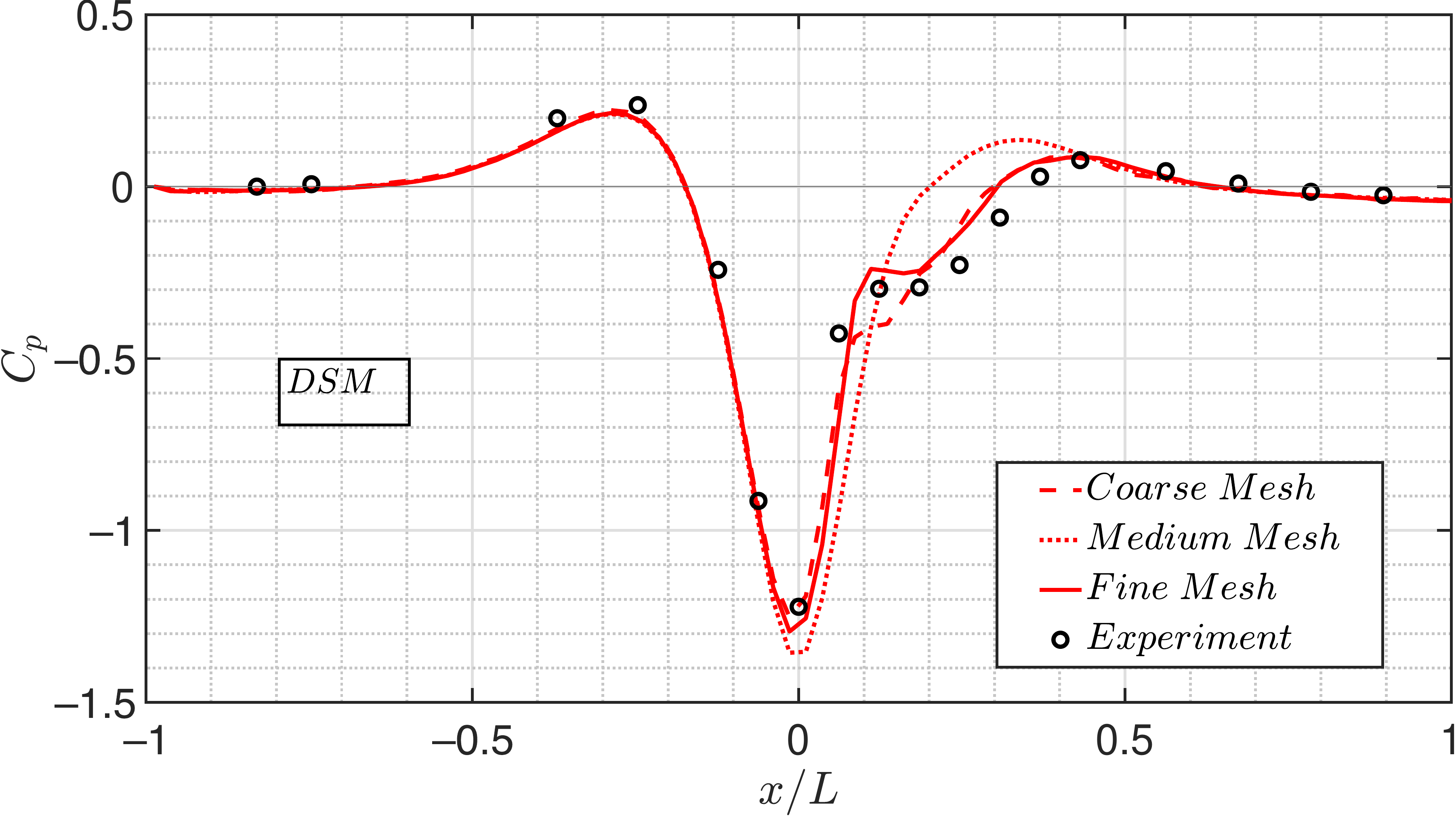}}
    \subfigure[]{
    \includegraphics[width=0.48\textwidth]{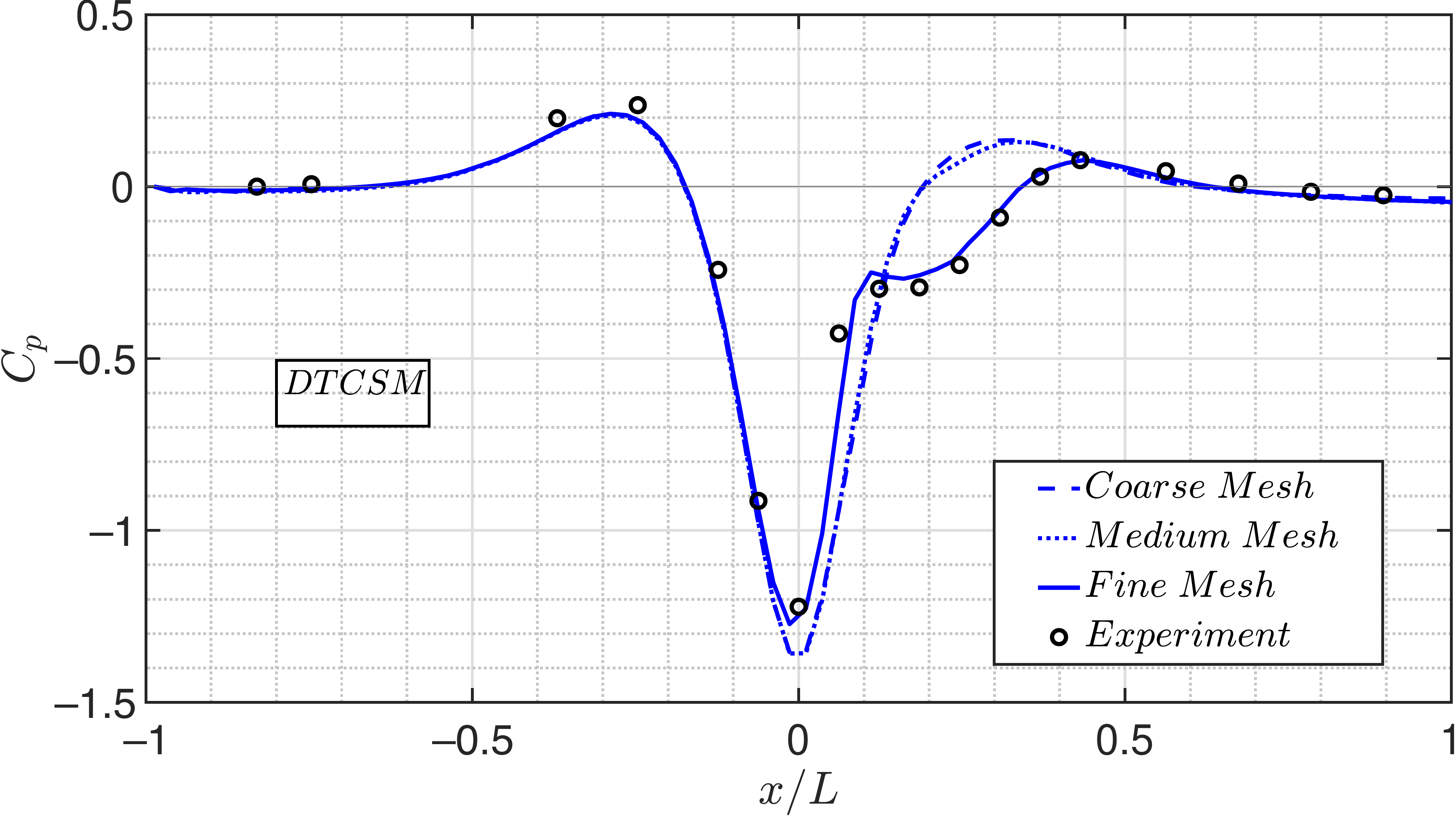}}
    \subfigure[]{
    \includegraphics[width=0.48\textwidth]{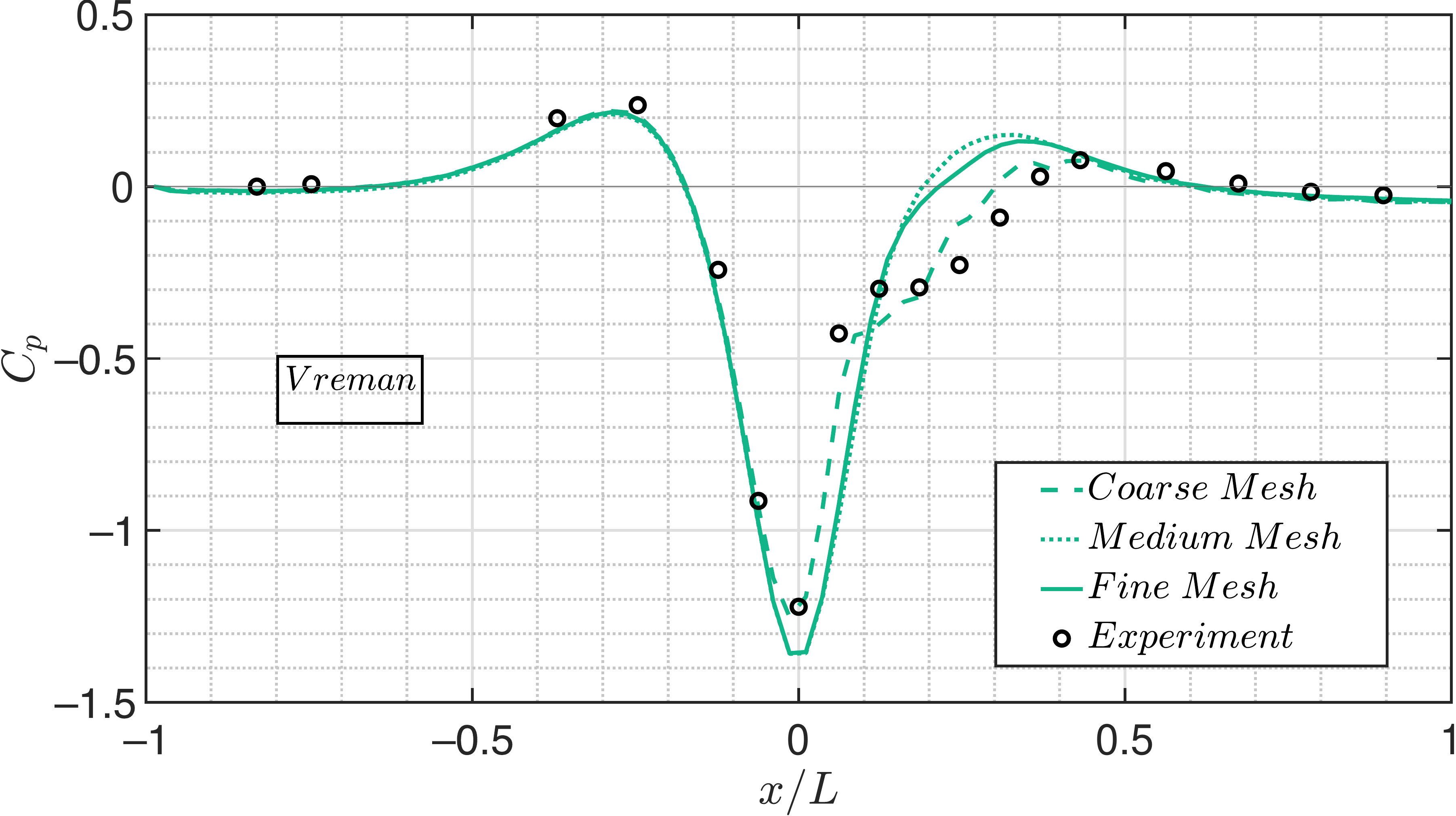}}
    \subfigure[]{
    \includegraphics[width=0.48\textwidth]{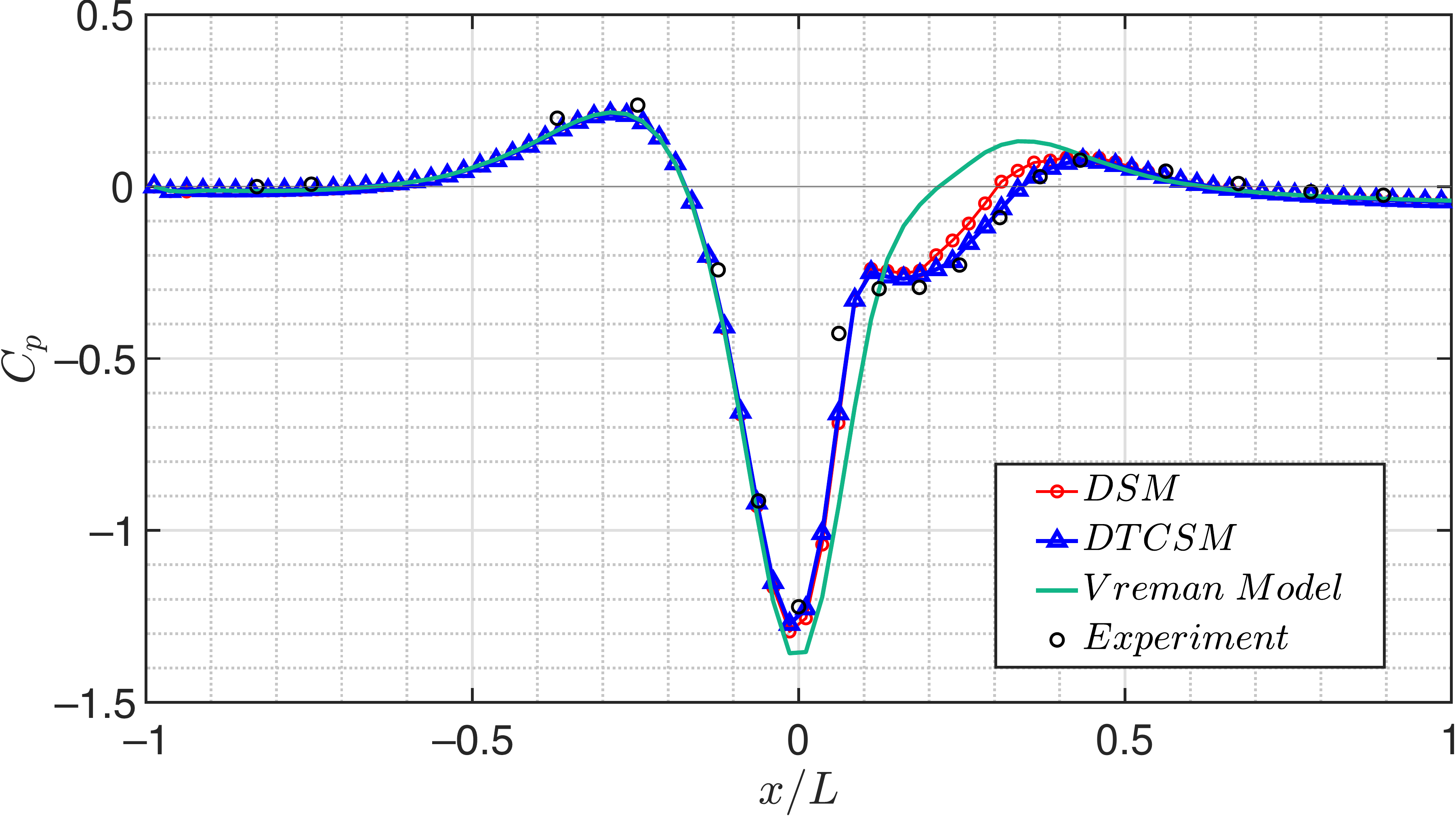}}
    \caption{  Streamwise distribution of the mid-span surface pressure coefficient for the three dimensional bump at $Re_L = 3.41 \times 10^6$ for (a) DSM, (b) DTCSM and (c)Vreman model. Finally, sub-plot (d) provides a comparison of the three models for the fine mesh. The black dots represent the experimental measurements of Williams et al. \cite{williams2020experimental}.}
    \label{fig:3dcp}
\end{figure}

The proposed tensor-coefficient model retains its monotonic convergence in the 3D configuration (and higher Reynolds number) unlike DSM. For DTCSM, as the grid is refined, the trough of $C_f$ moves towards DNS monotonically. The region of maximum flow acceleration ($x/L \sim -0.1$) where the skin friction reaches its peak is also more robust to refinement and more accurate for DTCSM. 
Figure \ref{fig:3dcp} compares the pressure coefficient at mid-span of the bump, where pressure profile flattens in the separation region. It becomes evident that the fine grid tensor-coefficient model produces excellent agreements with the experiments. The monotonicity of DTCSM also holds for the pressure unlike DSM in that for DTCSM the flow remains attached at coarse and medium grid levels. The suction peak is captured well by both DSM and DTCSM since it is primarily due to inviscid effects. Indeed, the suction peak is accurately predicted in a separate simulation (not shown) with free slip (inviscid) boundary conditions.

\begin{figure}[!ht]
    \centering
    \includegraphics[width=0.7\textwidth]{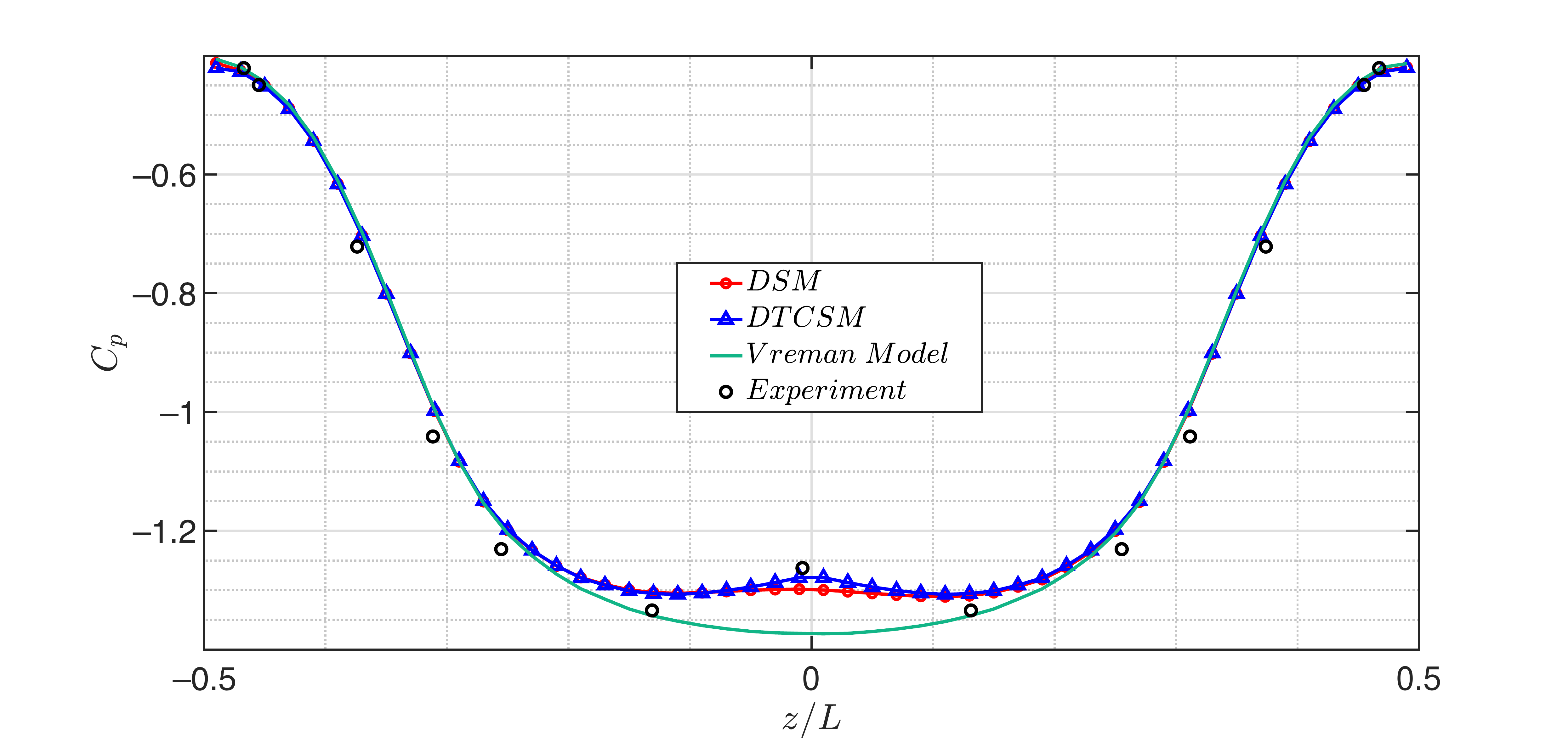}
    \caption{Spanwise distribution of the surface pressure coefficient for the three dimensional bump at $Re_L = 3.41 \times 10^6$ for DSM, DTCSM and Vreman model for the fine mesh case. The black dots represent the experiments of Williams et al. \cite{williams2020experimental}.}
    \label{fig:3dcps0}
\end{figure}

An important feature of this three dimensional flow, which is not-observed in the spanwise-periodic case is the variation in pressure across the span of the domain. Figure \ref{fig:3dcps0} compares the pressure coefficient variation at the apex of the bump for the fine grid cases. The slight rise in the pressure at mid-span is possibly due to the effect of the counter-rotating vortices that reduce the local streamwise velocity, which increases the local pressure in the region. Both models predict the pressure far away from the mid-span. Near the center, however, it is evident that \textcolor{black}{both DSM and} DTCSM capture the small and subtle rise in $C_p$, likely due to the improved effective body shape from an accurate prediction of the extent of the separation downstream. (Although not shown, our experiments have suggested that a larger separation bubble leads to a larger effective stream-wise extent of the bump, which shifts the value of $C_p$ at the suction peak towards zero). 

The relative under-performance of the Vreman model was established in the spanwise-periodic bump. In this three-dimensional flow, however, the differences become even more apparent in that the Vreman model does not predict separation even with the finest grids (refer to Figures \ref{fig:3dcf},\ref{fig:3dcp}). 
The issue with the Vreman model is also visible in the pressure prediction across the span at the bump apex where the pressure peak at mid-span is completely missed by the Vreman model (Figure \ref{fig:3dcps0}). 

\begin{figure}[!ht]
    \centering
    \subfigure[]{        \includegraphics[width=0.44\textwidth]{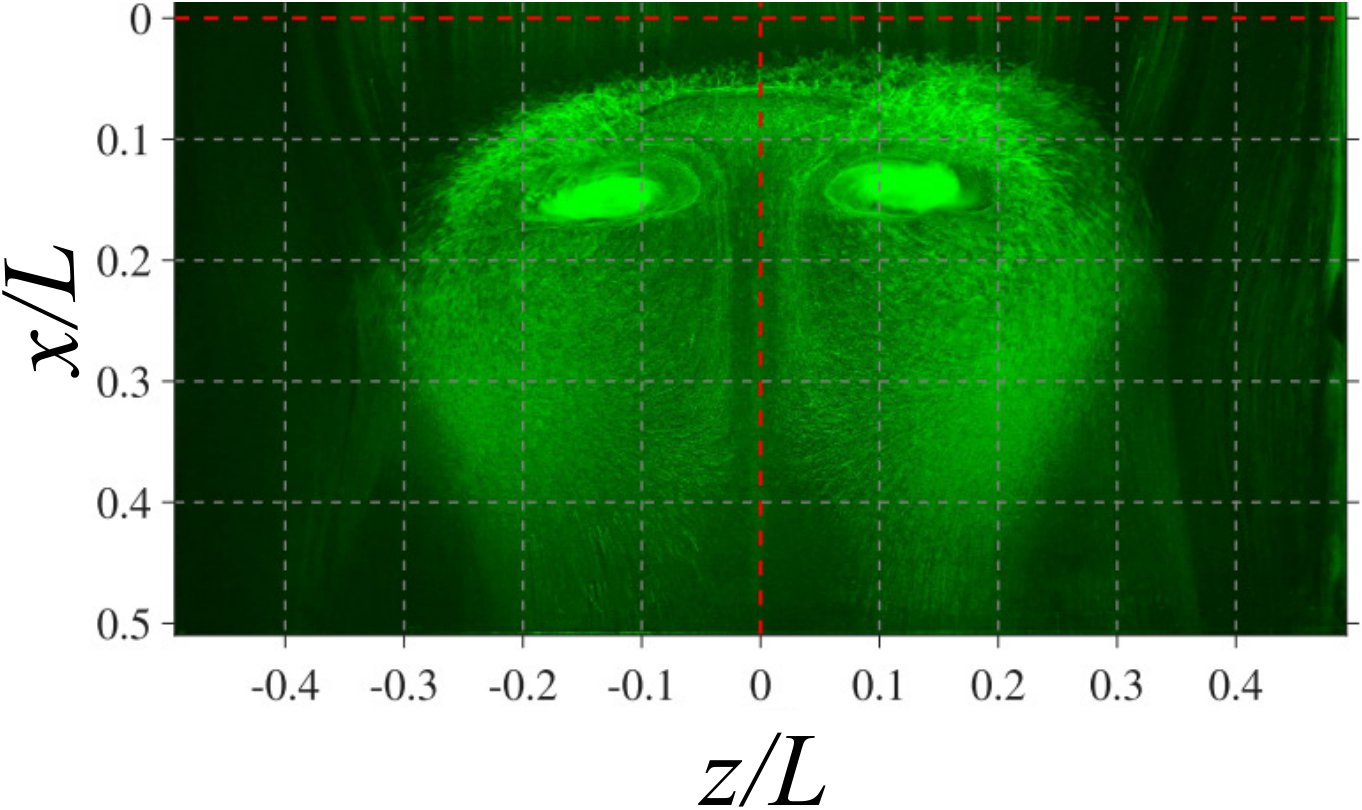} }
    \subfigure[]{
        \includegraphics[width=0.44\textwidth]{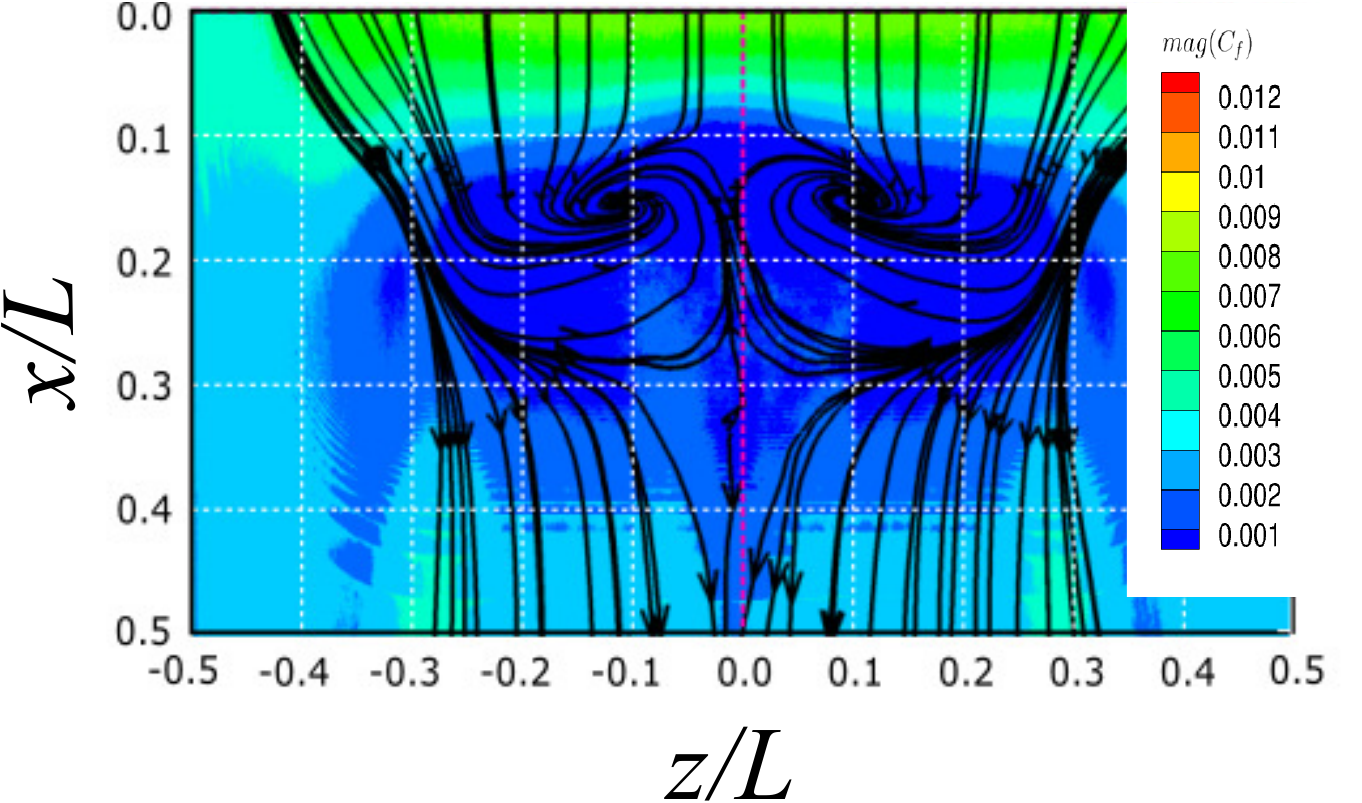}    }
    \subfigure[]{
        \includegraphics[width=0.44\textwidth]{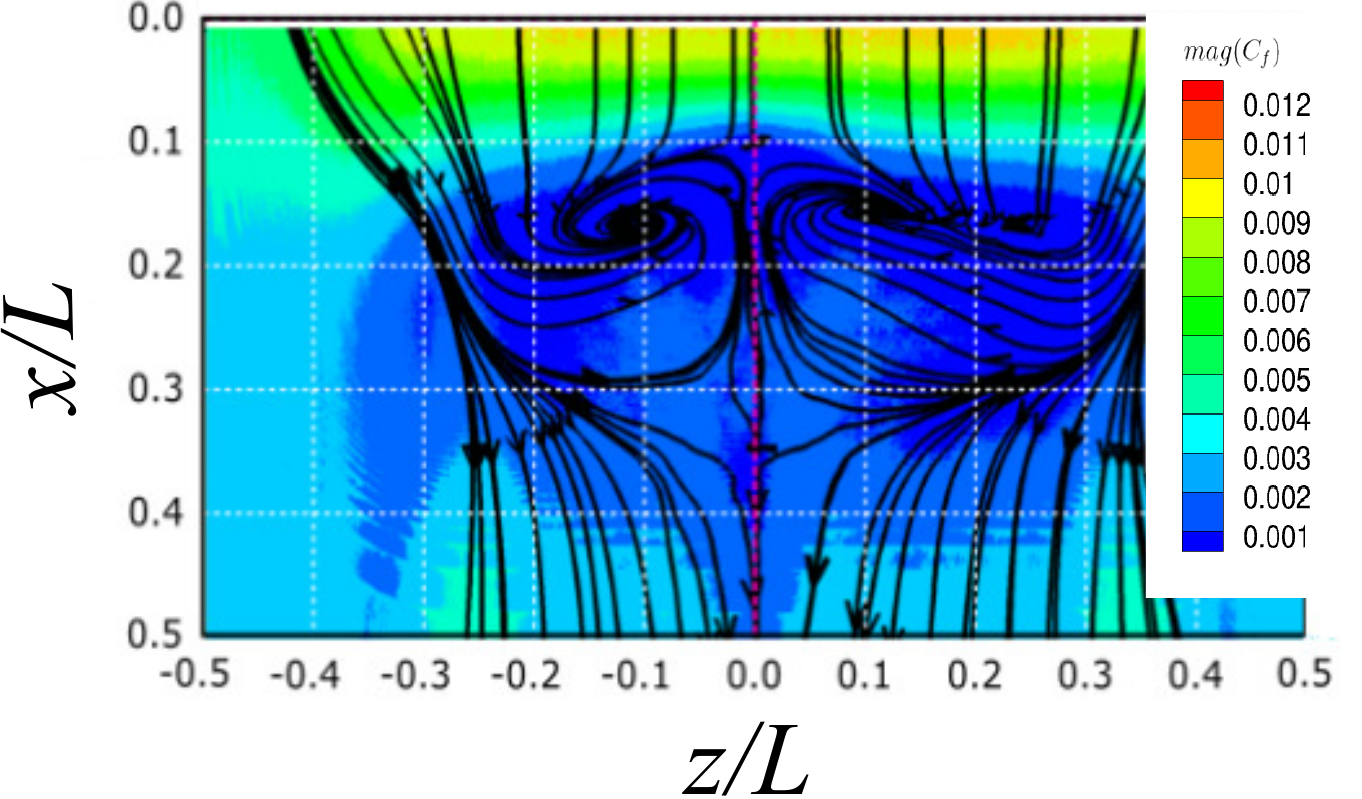}}
    \subfigure[]{        \includegraphics[width=0.44\textwidth]{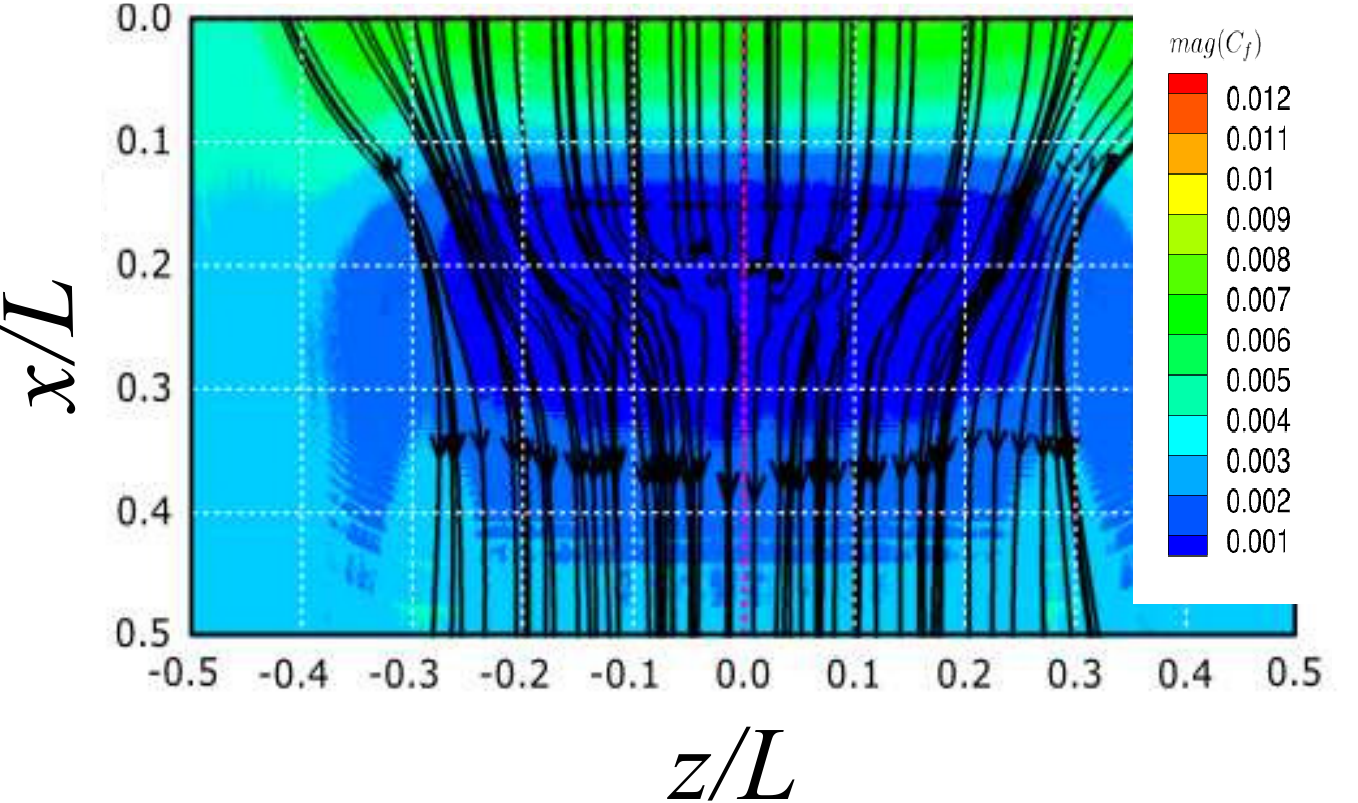}    }
    \caption{\textcolor{black}{(a) An oil-flow visualization of the counter-rotating vortices in the region of separation (as observed experimentally \cite{williams2020experimental}); reprinted with permission of the American Institute of Aeronautics and Astronautics, Inc. In sub-plots (b), (c) and (d) we compare the surface streamlines in the separation region for DSM, DTCSM and Vreman models respectively. It is also noticeable that for DTCSM, the skin friction at the apex of the bump is inline with the experimental value (from Figure \ref{fig:3dcf}) and higher than DSM and Vreman models.}}
    \label{fig:3dstrucexpt}
\end{figure}

\begin{figure}[!ht]
    \centering
    \subfigure[]{
        \includegraphics[width=0.7\textwidth]{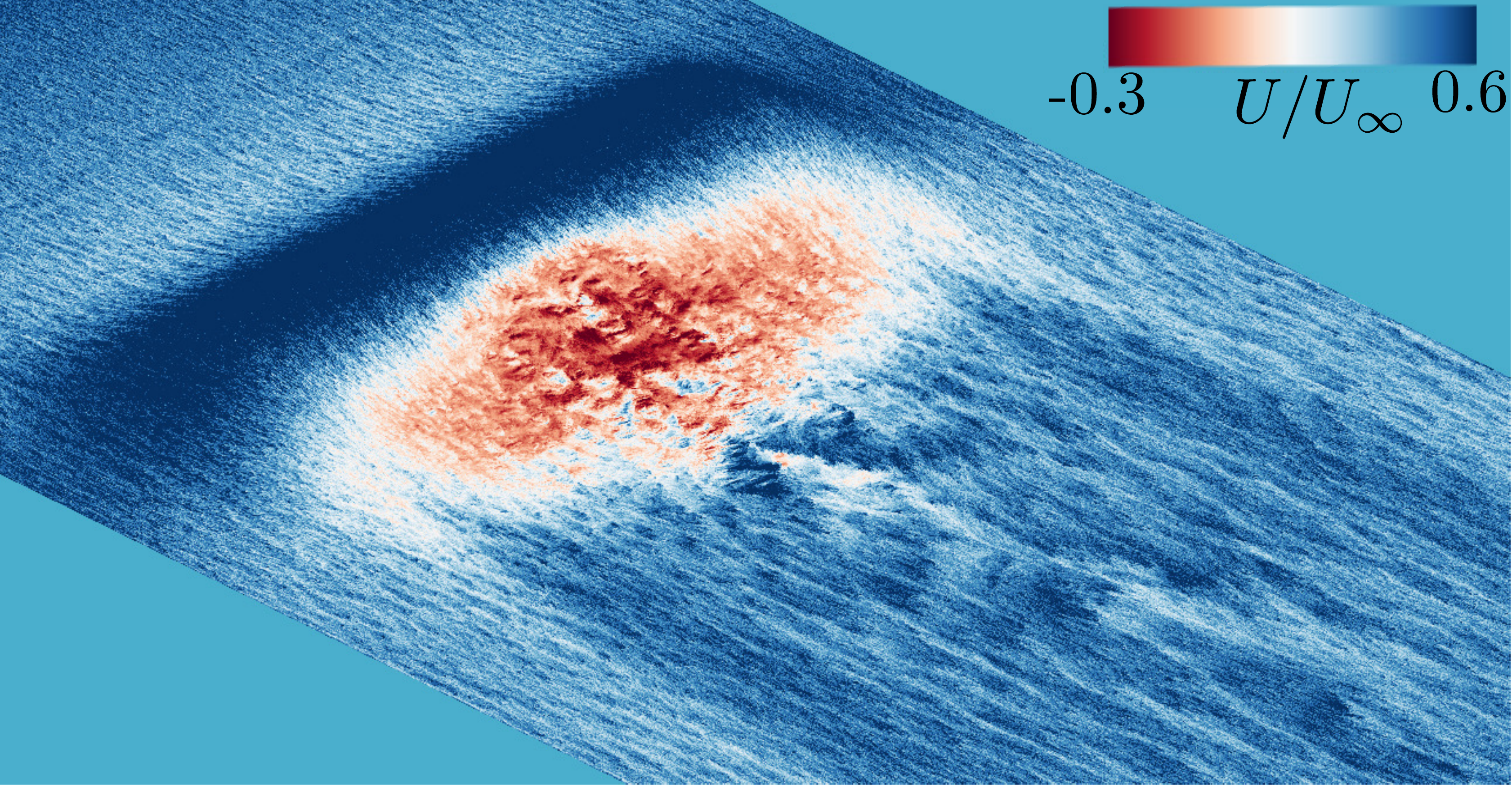}
    }
    \subfigure[]{
        \includegraphics[width=0.7\textwidth]{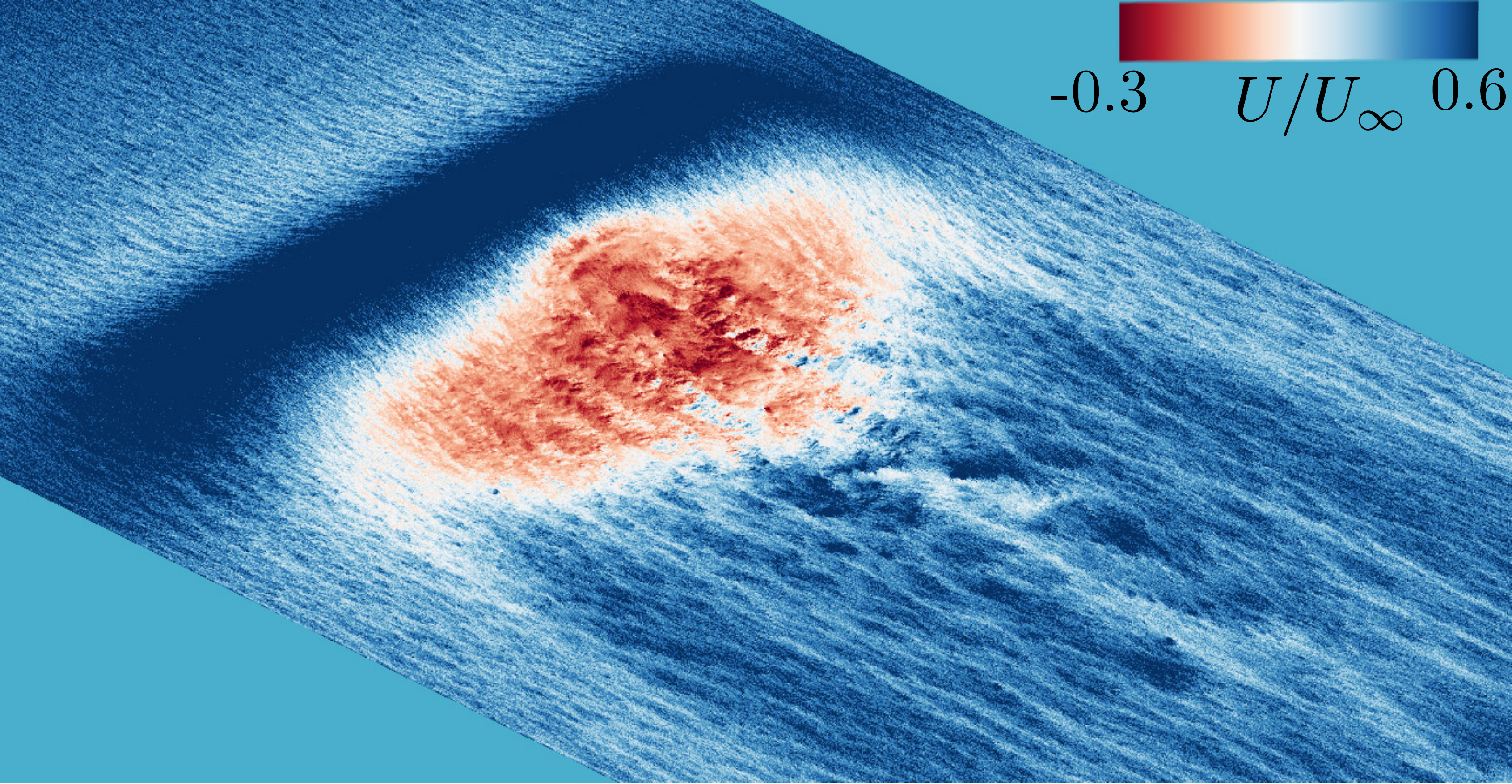}
    }
    \subfigure[]{
        \includegraphics[width=0.7\textwidth]{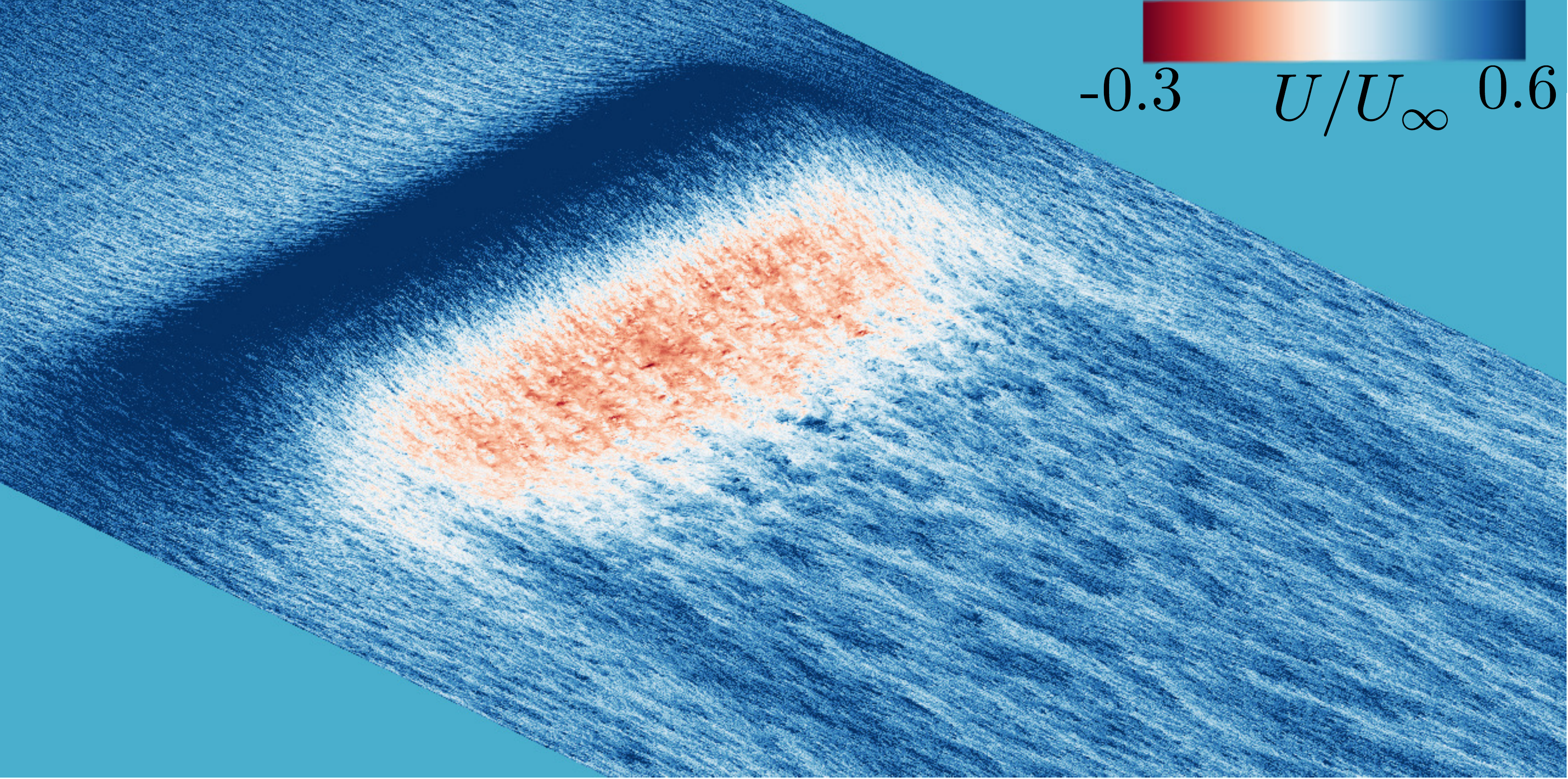}
    }
    \caption{\textcolor{black}{Oblique view of the instantaneous streamwise velocity in the first grid cell above the wall, projected on the bump surface for the fine mesh case. Sub-figures (a), (b), (c) are for DSM, DTCSM and Vreman models, respectively. }  }
    \label{fig:3dstrucsimx}
\end{figure}

Previously, comparisons between surface streamlines of skin friction from the LES have been compared with experimental visualizations to qualitatively study the nature of the flow structures \cite{goc2021large,goc2020wall,lozano2021performance}. For this flow, Williams et al. \cite{williams2020experimental} performed clay-kerosene mixture based visualizations and observed a pair of counter-rotating vortices in the aft section of the bump. In the current work, we compare the surface  skin-friction streamlines with these experimental visualizations. Our simulations with DSM and DTCSM accurately capture the location and size of the strongly vortical region as observed in figure \ref{fig:3dstrucexpt}. It is also evident that the Vreman model does not capture this vortex-pair, and thus predicts faster (and hence more attached) flow near the wall. The streamlines also reveal the characteristics of the mean circulation in the separated region. For DSM and DTCSM, it is clearly observed that the mean spanwise velocity is directed towards the center near the side walls, and an outward (from center) moving region is present near the mid-span. The Vreman model, on the contrary, does not predict this behavior owing to the absence of the separation bubble.  
This observation is consistent with the surface skin-friction streamlines of Iyer and Malik \cite{iyer2021wall} with the Vreman model. The instantaneous values of the streamwise velocity, projected on the wall from the nearest neighbouring cell  
for the fine mesh simulation are presented in Figure \ref{fig:3dstrucsimx}. Both DSM and DTCSM qualitatively behave similarly, in that the flow strongly decelerates (and eventually separates) on passing through the apex. Since the flow remains attached (in the mean) for the Vreman model, this deceleration is expected to be much weaker, and is confirmed in our calculations (Figure \ref{fig:3dstrucsimx}(c)). A vortex-wake extending up to approximately $x/L \sim 0.5$ is also observed for DSM and DTCSM, which is in agreement \textcolor{black}{with} the experimental observations. In this view, the recirculation zone appears to be the strongest for DTCSM, which is in agreement with its higher (and more accurate) $C_p$ prediction around mid-span. 

 \section{Remarks on Computational costs of performing LES using DTCSM }
\textcolor{black}{In this work, it has been observed that the computational cost of using DTCSM in LES is slightly higher than DSM by a factor of up to approximately 1.15, potentially due to the added requirement of dynamically evaluating multiple model coefficients.} 

\textcolor{black}{ Tables \ref{table:cost2D} and \ref{table:cost3D} describe the costs of simulating the Gaussian speed bump using the charLES flow solver for the finest grids considered in this work (Although not shown, the relative cost of performing simulations on coarser grids scale similarly for the two models). The non-dimensional time step (non-dimensionalized by the size of the domain in the streamwise direction, $L_x$ and the reference freestream velocity, $U_{\infty}$) is found to be approximately same for the two models. Further, the cost of advancing the governing equations up to one flow through time (defined as the ratio of $L_x$ and $U_{\infty}$) is only marginally higher for DTCSM.  Hence, it is apparent that the costs incurred on using DTCSM are not greatly different to using DSM for the same computational grid. }

 \begin{table}[!ht]
 \begin{tabular}{%
 >{\centering\arraybackslash}p{2.2cm}%
 >{\centering\arraybackslash}p{2cm}%
 >{\centering\arraybackslash}p{3cm}%
>{\centering\arraybackslash}p{3cm}%
 >{\centering\arraybackslash}p{3cm}%
 }
  SGS Model & Cell Count(Mil.) & Non-Dimensional time step & No. of GPU nodes (6 NVIDIA V100 cards per node) & GPU node hrs. per flow through time \\
  \hline\noalign{\vspace{3pt}}
  DSM  & 52 &  $4.74 \times 10^{-5}$ &10& 13 \\
  DTCSM &  52 &  $4.74 \times 10^{-5}$  &10& 15 \\
 \end{tabular}
  \caption{\textcolor{black}{Computational cost summary for the spanwise-periodic Gaussian bump using charLES flow solver per flow through time. }}
  \label{table:cost2D}
 \end{table}

 \begin{table}[!ht]
 \begin{tabular}{%
 >{\centering\arraybackslash}p{2.2cm}%
 >{\centering\arraybackslash}p{2cm}%
 >{\centering\arraybackslash}p{3cm}%
 >{\centering\arraybackslash}p{3cm}%
 >{\centering\arraybackslash}p{3cm}%
 }
  SGS Model & Cell Count(Mil.) & Non-Dimensional time step & No. of GPU nodes (6 NVIDIA V100 cards per node) & GPU node hrs. per flow through time \\
  \hline\noalign{\vspace{3pt}}
  DSM  & 452 &  $4.21 \times 10^{-5}$ &80& 100 \\
  DTCSM &  452 &  $4.21 \times 10^{-5}$  &80& 112 \\
 \end{tabular}
 
   \caption{ \textcolor{black}{Computational cost summary for the three dimensional Gaussian bump using charLES flow solver per flow through time.}}
  \label{table:cost3D}
 \end{table}
 
\section{Conclusions}

In this work, we have developed and validated a new non-Boussinesq-type dynamic SGS closure formulation for the tensor-coefficient Smagorinsky model (DTCSM). 
Similar to the dynamic Smagorinsky model, the only input parameter to this model is the ratio of test-level and grid-level filter widths. 

\emph{A priori} \textcolor{black}{analyses} have confirmed improved tensor-level correlations of exact SGS stresses and DTCSM. Asymptotic behavior of the model in the near-wall region is in-line with DNS data without the need for damping functions. The model also produces vanishing subgrid-scale stresses in laminar flow. Large-eddy simulations of decaying and forced \textcolor{black}{isotropic turbulence} at $\textit{Re}_{\lambda} = 70, 315$ and $Re_{\lambda} \rightarrow \infty $ have been performed. DTCSM performs well in HIT at all Reynolds numbers considered. For wall-modeled LES of a turbulent channel flow at $\textit{Re}_{\tau} = 4200$, improvements in mean-velocity profile, prediction of the K\'{a}rm\'{a}n constant and in mean streamwise, wall-normal intensities near the wall are observed.  Finally, it is \textcolor{black}{demonstrated} that DTCSM improves the prediction of the skin-friction peak for wall-modeled simulations of the flow over both the spanwise-periodic and three dimensional Gaussian bump at $Re_L = 2 \times 10^6 \; \mathrm{ and }\; 3.41 \times 10^6$ respectively and also does not suffer from non-monotonic convergence towards quasi-DNS unlike DSM. For the three-dimensional case, pressure in both the streamwise and spanwise directions are better predicted by DTCSM over DSM. The formation of the counter rotating vortex pair in the three dimensional bump is observed for the two dynamic subgrid-scale models in accordance with the experiment. \textcolor{black}{On the other hand, calculations with the constant coefficient Vreman model led to incorrect separation behavior. }\textcolor{black}{The computational cost incurred on performing LES with DTCSM is comparable (upto 15\% higher) to the cost of using the dynamic Smagorinsky model.} 



\section{Acknowledgments}
\textcolor{black}{This work was supported by the NASA’s Transformational Tools and Technologies project under grant number NNX15AU93A and by Boeing Research \& Technology. 
R.A. gratefully acknowledges support from the Stanford School of Engineering Fellowship.  K.P.G. acknowledges support from the Stanford Graduate Fellowship and the National Defense Science and Engineering Graduate Fellowship. We thank Ahmed Elnahhas for providing the DNS dataset of a turbulent channel flow at $Re_{\tau} = 395$. We also acknowledge helpful discussions with Konrad Goc. Computing resources were awarded through the Oak Ridge Leadership Computing Facility (DoE ALCC).  
}
 

\bibliographystyle{abbrv}
\bibliography{References}

\appendix

\section{Proof of tracelessness of DTCSM} \label{app:1}

In this appendix, we show that DTCSM is traceless with the constraints imposed in Eq. (\ref{eqn:dtcsm1}). This property is required for consistency as the quantity that is being modeled, the deviatoric part of the subgrid-scale stress tensor, is trace-free by definition. Recall that the model formulation of DTCSM is
\begin{equation}
\tau_{ij}^{sgs}  - \frac{\tau^{sgs}_{kk}  }{3} \delta_{ij}  = - (C_{ik}S_{kj} + C_{jk}S_{ki} )|S|\Delta^2 .
\label{eqn: dtcsmap}
\end{equation}
%
%
%
%
The trace of the right-hand side of Eq. \ref{eqn: dtcsmap} is 
\begin{equation}
    {\rm trace}\left[-(C_{ik}S_{ki} + C_{ik}S_{ki} )|S|\Delta^2\right] = -2 C_{ik}S_{ki}|S| \Delta^2~,
\end{equation}
which is further expanded as
\begin{equation}
    \begin{aligned} 
        -2C_{ik}S_{ki}|S|\Delta^2= &
        -2|S| \Delta^2 \left\{ C_{11} S_{11} + C_{22} S_{22} + C_{33}S_{33} + 
        S_{12}(C_{12}+C_{21})\right.\\ &\left. + S_{13}(C_{13}+C_{31}) +  S_{23}(C_{23}+C_{32})\right\}
    \end{aligned}
    ~.
\end{equation}
Assuming that the flow is incompressible ($S_{ii} = 0$) and that $S_{ij}$ is a general strain-rate tensor, we impose
\begin{equation}
C_{11} = C_{22} = C_{33} ~; \qquad
C_{ij} = - C_{ji} \quad (j \neq i) ~,
\end{equation}
which ensures that the subgrid stress predicted by DTCSM is trace-free.

\section{System of equations for dynamic procedure in DTCSM}  \label{app:2}

Recall for DTCSM, the Germano identity is written as
\begin{equation}
    L_{ij} = ( C_{ik} \Delta^2 M_{kj} + C_{jk} \Delta^2 M_{ki} ) ~.
\end{equation}
Absorbing the factor of $\Delta^2$ into the coefficients $C_{ij}$, and including only the free coefficients of $C_{ij}$, the system can be rewritten as
\begin{equation}
    \begin{pmatrix}
        L_{11} \\
        L_{22} \\
        L_{33} \\
        L_{12} \\
        L_{13} \\
        L_{23} \\ 
    \end{pmatrix} =    \begin{pmatrix}
        2 M_{11} & 2 M_{12} & 2 M_{13} & 0  \\
        2 M_{22} & -2 M_{12} & 0 & 2 M_{23}  \\
        2 M_{33} & 0 & - 2 M_{13} & -2 M_{23}  \\
        2 M_{12} & M_{22} - M_{11} & M_{23} & M_{13}  \\
        2 M_{13} & M_{23}  & M_{33}-M_{11} & -M_{12}  \\
        2 M_{23} & -M_{13} & -M_{12} & M_{33}- M_{22}  \\
    \end{pmatrix}       \begin{pmatrix}
        C_{11} \\
        C_{12} \\
        C_{13} \\
        C_{23} \\
    \end{pmatrix}~, 
\end{equation}
which, by defining the $6\times4$ matrix on the right-hand side as $M_{mat}$, and $\left[\cdot\right]$ as \textcolor{black}{vector}, can be written concisely as
\begin{equation}
    [L] = M_{mat}[C]~.
\end{equation}
Using the least-squares solutions approach, we get
\begin{equation}
    M_{mat}^T [L] = M_{mat}^T  M_{mat}[C]~.
    \label{eqn:dtcsmfinal}
\end{equation}
Finally, Eq. (\ref{eqn:dtcsmfinal}) is solved directly to dynamically evaluate the four model coefficients.
%
%
%
%
%

\section{Another representation of DTCSM }  \label{app:3}

In this appendix, we re-express the model form of DTCSM in terms of a combination of the strain-rate and rotation-rate tensor. In general, the stresses from DTCSM are expressed as 
\begin{equation}
        \tau^{DTCSM}_{ij} = -( C_{ik}  S_{kj} + C_{jk} S_{ki}) |S| \Delta^2.
\end{equation}
%
Using the realizability constraints on DTCSM, and decomposing the model coefficient matrix $C_{ij}$ into isotropic and deviatoric parts, $C_{ij} = C_{11} \delta_{ij} + C^d_{ij}$, we write
\begin{equation}
    \begin{aligned}
        \tau_{ij}^{DTCSM} = &
        - \left\{ ( C_{11} \delta_{ik}+ C^d_{ik} )S_{kj} + ( C_{11} \delta_{jk}  + C^d_{jk} ) S_{ki} \right\}|S| \Delta^2 ~.
    \end{aligned}
    \label{eqn:716}
\end{equation}
Since the deviatoric part of the coefficient matrix $C^d_{ij}$ is constrained to be anti-symmetric, it therefore has properties similar to those of the rotation-rate tensor. Setting $|S| C_{ij}^d = - \Lambda  R_{ij}$, Eq. (\ref{eqn:716}), the model can be simplified to the form 
\begin{equation}
     \tau_{ij}^{DTCSM}  =
    - 2 C_{11}\Delta^2 S_{ij} |S|
    - \Lambda \Delta^2  \left( S_{ik} R_{kj} - R_{ik} S_{kj} \right ) ~,
\end{equation}
which is the same model form as the two-term expansion of the velocity gradient tensor in the previous work of Lund and Novikov \cite{lund1992parameterization} and a recent study by Agrawal et al. \cite{agrawalarb2021}. Hence, under certain scenarios, DTCSM can be interpreted as a model which explicitly accounts for effect of large-scale rotation rates unlike DSM. 

\end{document}